\documentclass[aps,prb,twocolumn,amsmath,amssymb,longbibliography,balancelastpage,citeautoscript,eqsecnum,10pt]{revtex4-1}

\pdfoutput=1

\usepackage{xcolor,mathtools,bm,mathrsfs,dsfont}
\usepackage[english]{babel}
\usepackage[babel,kerning=true,spacing=true]{microtype}

\definecolor{darkblue}{RGB}{0,0,150}
\definecolor{nightblue}{RGB}{0,0,100}

\usepackage[
colorlinks,
citecolor=darkblue,
linkcolor=darkblue,
urlcolor=nightblue]{hyperref}

\bibpunct{[}{]}{,}{n}{}{}
\makeatletter
\def\NAT@def@citea{\def\@citea{\NAT@separator}}
\makeatother

\let\Re\relax
\let\Im\relax
\DeclareMathOperator{\Re}{Re}
\DeclareMathOperator{\Im}{Im}
\DeclareMathOperator{\Tr}{Tr}

\newcommand{\dd}{\mathrm{d}}

\newcommand{\annrevcond}{Ann. Rev. Cond. Mat.}
\newcommand{\arx}{}

\newcommand{\prx}{Phys. Rev. X}
\newcommand{\njp}{New J. Phys.}

\begin{document}

\title{Electronic properties of disordered Weyl semimetals at charge neutrality}

\date{{\normalsize{}\today}}

\author{Tobias Holder$^{1,2}$, Chia-Wei Huang$^{1}$, and Pavel Ostrovsky$^{1}$}

\affiliation{${}^1$Max Planck Institute for Solid State Research, Heisenbergstrasse
1, 70569, Stuttgart, Germany\\
${}^2$Department of Condensed Matter Physics, Weizmann Institute of Science, Rehovot, Israel 76100}
\begin{abstract}
Weyl semimetals have been intensely studied as a three dimensional realization of a Dirac-like excitation spectrum where the conduction bands and valence bands touch at isolated Weyl points in momentum space. Like in graphene, this property entails various peculiar electronic properties. However, recent theoretical studies have suggested that resonant scattering from rare regions can give rise to a non-zero density of states even at charge neutrality. Here, we give a detailed account of this effect and demonstrate how the semimetallic nature is suppressed at the lowest scales. To this end, we develop a self-consistent T-matrix approach to investigate the density of states beyond the limit of weak disorder. Our results show a nonvanishing density of states at the Weyl point which exhibits a non-analytic dependence on the impurity density. This unusually strong effect of rare regions leads to a revised estimate for the conductivity close to the Weyl point and emphasizes possible deviations from semimetallic behavior in dirty Weyl semimetals at charge neutrality even with very low impurity concentration.
\end{abstract}
\maketitle

\section{Introduction}

The last few years saw significant progress in the study of gapless systems with non-trivial topology, mainly fueled by the successful generalization of the well established phenomenology of topological insulators (TI) to gapless systems. 
In a TI, where the band structure of an insulator supports a topological non-trivial state, the bulk remains gapped, whereas counterpropagating edge states are present on the surface~\cite{Hasan2011,Qi2011}. 
Topological phases are however not constrained to materials with a bulk gap, 
gapless topological phases can be realized in systems with protected Fermi points and superconductors with nodes~\cite{Matsuura2013}. 
In these materials, the simultaneous presence of non-trivially realized global symmetries and gapless, most often relativistic fermions offers a fertile ground for numerous exotic quantum phases~\cite{Turner2013,Vafek2014,Metlitski2014}.

A well-studied type of gapless topological system is the Weyl semimetal (WSM)~\cite{Burkov2011,Wan2011}, which features an even number of non-degenerate Fermi surface points with linear Dirac cone dispersion. In these systems, the lifting of degeneracy is facilitated by breaking of time-reversal symmetry. The chiral Weyl cones are sources and sinks of Berry flux, leading to non-trivial topological properties.

WSM exhibit a rich phenomenology, including the chiral anomaly~\cite{Nielsen1983,Son2012, Chen2013}, a large negative magnetoresistance~\cite{Son2013,Burkov2014,Gorbar2014}, as well as special protected surface states (Fermi arcs)~\cite{Wan2011,Baum2015}. 
Experimentally, TaAs~\cite{Xu2015,Lv2015} and NbAs~\cite{Xu2015a} have been found to harbor a WSM, 
some experiments have also seen signatures of Fermi arcs~\cite{Xu2015a,Lv2015a} and the detection of the chiral anomaly has been reported~\cite{Zhang2016}.

Quite some attention was attributed to the effects of disorder in a WSM. 
Since short-ranged impurities are irrelevant in three dimensions, an instability towards a diffusive metal is only expected at a finite impurity density~\cite{Fradkin1986,Fradkin1986a}.
The associated quantum critical point has been studied intensely, both with respect to critical scaling laws~\cite{Goswami2011,Roy2014,Syzranov2015,Sbierski2015,Liu2016,Syzranov2016} and the wider phase diagram~\cite{Sbierski2014,Ominato2014,Kobayashi2014,Pixley2015,Pixley2016,Bera2016,Roy2016}.
In the limit of weak disorder, the calculation of the dc-conductivity was done early on~\cite{Burkov2011,Burkov2011a,Hosur2012}, later the focus shifted to the exploration of magnetoconductance~\cite{Son2013,Burkov2014,Gorbar2014,Burkov2015,Klier2015,Altland2015,Ramakrishnan2015,Khalaf2016}.

Directly at charge neutrality (chemical potential $E=0$), the density of states (DOS) vanishes and with it the effects of any weak disorder. 
A popular choice to model this situation is a Gaussian distribution for the disorder, which in turn allows to calculate the self-energy and thus the scattering rate $\Gamma$ with the help of the self-consistent Born approximation (SCBA)~\cite{Burkov2011,Hosur2012,Ominato2014,Sbierski2014,Tabert2016}.
For the dc-conductivity, the Boltzmann approach then yields $\sigma_{dc}\sim E^2/\Gamma$~\cite{Burkov2011,Hosur2012,Lundgren2014}. Together with the quadratic energy dependence of the DOS ($\rho\sim E^2\sim\Gamma$) this leads to a nonzero conductivity which diverges with decreasing disorder.
With the Kubo formalism, the conductivity contains an additional term proportional to the scattering rate~\cite{Ominato2014,Tabert2016}. It can be shown that irrespective of the precise form of $\Gamma(E)$, a generic expression holds for $\sigma_{dc}$ in the regime of weak disorder~\cite{Tabert2016}, 
\begin{align}
\sigma_{dc}&=\sigma_0\frac{E^2+3\Gamma(E)^2}{\Gamma(E)}.
\label{eq:genericdccond}
\end{align}
A few studies have also elaborated on the $\omega$ and T-dependence of the conductivity within SCBA~\cite{Burkov2011,Hosur2012,Rosenstein2013,Ashby2014,Tabert2016}.

This simple picture of weak disorder does however not differentiate between a small impurity density $n_i$ and weak impurities with a small scattering potential $V$~\cite{Nandkishore2014}. In particular, it cannot account for rare cases where the Weyl fermions are repeatedly (resonantly) scattered.
Recently another issue was pointed out~\cite{Biswas2014}. The WSM features an anisotropic scattering amplitude, which results in a difference between the quasiparticle lifetime and transport lifetime. This is due to the intrinsically relativistic dispersion, leading to the interlocking of spin with rotational degrees of freedom. This property affects the coefficients of various response functions, but was not considered in the initial treatments of disordered WSMs. Here, we address the problem of resonant scattering within a microscopic calculation, which fully accounts for the mentioned idiosyncracies of Weyl semimetals.

As a side remark, we point out that charged impurities with a long-ranged impurity potential lead to an entirely different situation. This kind of disorder is  relevant and leads to the formation of charge puddles~\cite{Skinner2014}. Therefore, even for $E=0$, a zero DOS is avoided and a minimal dc-conductivity proportional to $n_i^{1/3}$ is found for the low temperature limit~\cite{Skinner2014,Ominato2015,Rodionov2015}. However, charged impurities themselves shift the chemical potential away from $E=0$, making it difficult to tune a WSM with dopants to the degeneracy point.

The focus of this work are the effects of resonant scattering, which appears in the presence of strong but dilute short-ranged impurities. In this case, if the impurity potential is tuned to a resonance, the total scattering cross section of an impurity center diverges. 
This behavior, also termed ``rare-region effect'' is the leading contribution to electronic transport close to the Weyl point and was identified as a way to avoid the quantum critical point between WSM and diffusive metal~\cite{Nandkishore2014,Pixley2016a}. 
In other words, in the presence of short-ranged disorder, even the perfectly tuned WSM does not retain a vanishing density of states (DOS) down to the smallest energies. As short-ranged impurities are irrelevant, this effect is non-perturbative in disorder strength and is not visible within the SCBA. In the analogous two-dimensional case of graphene, disorder is marginal and a finite DOS is recovered already from the SCBA~\cite{Ostrovsky2006}.

The importance of rare regions for WSMs was first pointed out by \citeauthor{Nandkishore2014}~\cite{Nandkishore2014} by demonstrating the existence of power-law bound states. In particular, they derived the phase shift of the associated scattering problem and obtained a exponentially small DOS for the case of a short-range correlated disorder distribution with zero mean.
A more recent numerical study by \citeauthor{Pixley2016a}~\cite{Pixley2016a} investigated the effect of resonant scattering using a Gaussian disorder model in a finite system. They concluded that a finite DOS is indeed recovered in the thermodynamic limit but might be very small. 

While the importance of resonant scattering is established for a WSM tuned to the Weyl point, the corresponding calculation of the associated self-energy, DOS and conductivity has not been done. 
This extends to the question of length scales and the universality of the results, both of which was only sketched in~\cite{Nandkishore2014}.
On a more technical level, just like for weak impurities the scattering amplitude is anisotropic, and the conductivity is expected to be renormalized by the contribution of particle hole ladders. 
This makes it necessary to correctly capture the diffusion pole by using an approximation which preserves charge conservation.

Here, we calculate self-energy, DOS and dc-conductivity in the presence of strong impurities for $T=0$ by solving the scattering problem of a rectangular impurity potential. Central in the derivation is the determination of the self-consistent self-energy and scattering T-matrix, which we determine in accordance with the Ward identities.
Preserving the diffusion pole, a universal form of the dc-conductivity emerges, which spans the entire parameter regime from weak to strong impurity centers.

As the main results, we show that scattering from rare regions leads to a self-energy proportional to the square root of the impurity density $n_i$, in sharp contrast to the perturbative regime given by SCBA. This affects the estimates for the energy windows in which a WSM crosses over from effectively weak to strong scattering, resulting in significantly different predictions for the mean free path close to the Weyl point.

The remaining sections are structured as follows. Section~\ref{sec:II} contains a brief exposition of the microscopic scattering problem of Weyl fermions hitting an impurity with a rectangular potential barrier. Then, in section~\ref{sec:III}, the self-energy is determined within the self-consistent T-matrix approach. For this, we resolve both the energy and momentum dependence of the scattering T-matrix close to resonance. The calculation of DOS and dc-conductivity are presented in~\ref{sec:IV}, focusing on the impact of the self-consistent self-energy and of the renormalized current vertex. We conclude in section~\ref{sec:V}.

\section{Weyl fermions scattered by a potential barrier}

\label{sec:II}

In this section we derive the scattering cross section of a Weyl fermion
encountering a spherical potential energy barrier.
The system of our interest is defined in a spherical coordinate with the radial vector $\bm{r}$, 
polar angle $\theta,$ 
and azimuthal angle $\varphi$. 
We consider a Weyl fermion of energy $E$ incident in the direction of $\mathbf{v}$ (angles $\zeta$, $\eta$) and scattered by a spherically symmetric potential of strength $\mathcal{V}(r)$, where $\mathcal{V}(r)$ has
a nonzero value $V$ for $r<b$, and $0$ otherwise. 
The direction of the scattered wave is labeled as $\mathbf{v'}$,
with energy $E'$ and angles $\zeta'$, $\eta'$. 
While we treat the single scattering of a plane wave which is hitting the impurity, the direction of the outgoing wave $\mathbf{v'}$ is equal to the direction of observation $\hat{\bm{r}}=\frac{\bm{r}}{r}$. For elastic scattering it is additionally $E=E'$, later on in section~\ref{sec:III} intermediate states will be taken into account.

\subsection{Helicity eigenbasis}

The scattered states can be obtained by solving the Schr\"odinger equation
with the Hamilton operator 
\begin{align}
H & =v\bm{\sigma}\cdot\bm{p}+\mathcal{V}(r)\label{eq:hami}
\end{align}
where $\bm{\sigma}$ is the vector of Pauli matrices. Henceforth we
will use the convention that $\hbar v=1$ and only restore units when
necessary.

Eigenstates of a clean system can be represented as plane waves, which are normally labeled by the momentum vector $\bm{k}$ and the chirality parameter $\lambda=\pm1$ (corresponding to the projection of spin onto the momentum direction). 
The wave function of such a state can be written as 
\begin{align}
|\bm{k},\lambda\rangle & =e^{i\bm{k}\cdot\bm{r}}|\lambda\hat{\bm{k}}\rangle\shortintertext{where}|\bm{n}\rangle & =\left(\begin{array}{c}
\cos(\theta/2)\\
\sin(\theta/2)e^{i\phi}
\end{array}\right)
\end{align}
The ket notation is used for the spinor in the direction $\bm{n}$ with the polar and azimuthal angles $\theta$ and $\phi$. 
The eigenstate energy is $E=\lambda|\bm{k}|$. 
Instead of using the momentum $\bm{k}$ and chirality $\lambda$ to label states, 
here we use an alternative representation in terms of energy $E$ and a unit velocity vector $\mathbf{v}$ as follows 
\begin{align}
|E,\mathbf{v}\rangle & =e^{iE\mathbf{v}\cdot\bm{r}}|\mathbf{v}\rangle.\label{eq:helicitybase}
\end{align}
This representation is known as helicity basis in the study of relativistic
systems, and it is related to the momentum representation by $\bm{k}=E\mathbf{v}$
and $\lambda=\mathrm{sgn}\, E$. The normalization has the form 
\begin{align}
\langle E,\mathbf{v}|E',\mathbf{v'}\rangle & =\frac{(2\pi)^{3}}{E^{2}}\delta(E-E')\delta(\mathbf{v}-\mathbf{v'}).
\end{align}

As the clean Weyl Hamiltonian $H=v\bm{\sigma}\cdot\bm{p}$ conserves the total angular momentum $\bm{J}$, one can also construct eigenstates using the total angular momentum number $j$ and the z-axis component $j_{z}$ as quantum numbers, yielding spherical waves 
\begin{align}
|E,j,j_{z}\rangle^{(1,2)} & =\frac{E}{\sqrt{2}}\biggl[h_{j-1/2}^{(1,2)}(Er)|\phi_{j,j_{z}}^{+}\rangle\nonumber \\
 & \qquad+ih_{j+1/2}^{(1,2)}(Er)|\phi_{j,j_{z}}^{-}\rangle\biggr].
\end{align}
Here, $h_{l}^{(1,2)}(x)$ are the spherical Hankel functions and $|\phi_{j,j_{z}}^{\pm}\rangle$ are spherical spinors. 
The index $(1,2)$ corresponds to waves which propagate outwards or inwards form the origin, respectively. The normalization is chosen such that the total particle flux in these states equals unity. 
Details of the spherical wave basis $|\phi_{j,j_{z}}^{\pm}\rangle$,
the construction of an eigenbasis for $H$ in the clean system and the partial wave expansion for the scattering problem can be found in appendix~\ref{sec:appA}. 
In the following, we focus on the solution for a rectangular impurity potential, $\mathcal{V}(r)=\Theta(b-r)V$.

\subsection{Scattering at a rectangular potential}

\label{sec:partialwave}

For the eigenstates of the Hamiltonian Eq.~(\ref{eq:hami}) we make
the ansatz

\begin{align}
|\psi_{Ejj_{z}}\rangle & =f(r)|\phi_{j,j_{z}}^{+}\rangle-ig(r)|\phi_{j,j_{z}}^{-}\rangle,
\end{align}
which yields two coupled differential equations for the radial functions
$f$ and $g$. The solutions are given in terms of spherical Bessel
functions of the first and the second kind, $\mathscr{\textrm{j}}_{l}(x)$
and $\mathscr{\textrm{n}}_{l}(x)$. Constructing the total wave function
of the scattering problem $|\psi\rangle$ as a sum over all angular
parts $|\phi_{j,j_{z}}^{\pm}\rangle$, one obtains in presence of
the impurity \looseness=-1

\begin{widetext}
\begin{equation}
|\psi\rangle=\begin{cases}
\sum_{jj_{z}}A_{jj_{z}}\left[\mathscr{\textrm{j}}_{j-1/2}(\bar{E}r)|\phi_{j,j_{z}}^{+}\rangle-i\mathscr{\textrm{j}}_{j+1/2}(\bar{E}r)|\phi_{j,j_{z}}^{-}\rangle\right]\;, & r<b\\
\sum_{jj_{z}}A'_{jj_{z}}\left[\mathscr{\textrm{j}}_{j-1/2}(Er)|\phi_{j,j_{z}}^{+}\rangle-i\mathscr{\textrm{j}}_{j+1/2}(Er)|\phi_{j,j_{z}}^{-}\rangle\right]\;\\
\quad+B'_{jj_{z}}\left[\mathscr{\textrm{n}}_{j-1/2}(Er)|\phi_{j,j_{z}}^{+}\rangle-i\mathscr{\textrm{n}}_{j+1/2}(Er)|\phi_{j,j_{z}}^{-}\rangle\right], & r>b,
\end{cases}\label{eq:totalwavefunction}
\end{equation}
\end{widetext}

where we introduced $\bar{E}=E-V$. We point out that $|\psi\rangle$ must
be regular at the origin, therefore there is no coefficient $B_{jj_{z}}$
for $r<b$. The coefficients $A_{jj_{z}}'$ and $B_{jj_{z}}'$ are
related to $A_{jj_{z}}$ by matching the wave functions at $r=b$,
yielding 
\begin{align}
A_{jj_{z}}' & =-A_{jj_{z}}N_{j}(\bar{E},E)\left(Eb\right)^{2},\label{eq:A'}\\
B_{jj_{z}}' & =A_{jj_{z}}M_{j}(\bar{E},E)\left(Eb\right)^{2},\label{eq:B'}
\end{align}
where we have used the fact that $\mathscr{\textrm{j}}_{n+1}\left(x\right)\mathscr{\textrm{n}}_{n}\left(x\right)-\mathscr{\textrm{j}}_{n}\left(x\right)\mathscr{\textrm{n}}_{n+1}\left(x\right)=x^{-2},$
and defined two shorthands 
\begin{align}
N_{j}(\bar{E},E) & =\mathscr{\textrm{j}}_{j-1/2}(\bar{E}b)\mathscr{\textrm{n}}_{j+1/2}(Eb)\nonumber \\
 & \quad-\mathscr{\textrm{n}}_{j-1/2}(Eb)\mathscr{\textrm{j}}_{j+1/2}(\bar{E}b),\label{eq:N}\\
M_{j}(\bar{E},E) & =\mathscr{\textrm{j}}_{j-1/2}(\bar{E}b)\mathscr{\textrm{j}}_{j+1/2}(Eb)\nonumber \\
 & \quad-\mathscr{\textrm{j}}_{j-1/2}(Eb)\mathscr{\textrm{j}}_{j+1/2}(\bar{E}b).\label{eq:M}
\end{align}

Scattering means that the incident plane wave is disturbed so that
a part of the current is redistributed to other directions $\hat{\bm{r}}\neq\mathbf{v}$,
i.~e. the total wave function can be written as 
\begin{align}
|\psi\rangle & =e^{iE\mathbf{v}\cdot\bm{r}}|\mathbf{v}\rangle+f(\mathbf{v},\hat{\bm{r}})\frac{e^{iEr}}{r}|\hat{\bm{r}}\rangle.\label{eq:scatteringwf}
\end{align}
Here, $f(\mathbf{v},\bm{r})$ is the amplitude for the scattering
of the incident wave from $(\cos\frac{\zeta}{2},\sin\frac{\zeta}{2}e^{i\eta})$
basis to $(\cos\frac{\theta}{2},\sin\frac{\theta}{2}e^{i\varphi})$
basis. Matching the coefficients of Eq.~(\ref{eq:totalwavefunction}) to the form of Eq.~(\ref{eq:scatteringwf}) for $r\rightarrow\infty$, one obtains 
\begin{align}
f(\mathbf{v},\bm{r}) & =\frac{4\pi}{iE}\sum_{jj_{z}}\biggl[\left\langle \phi_{j,j_{z}}^{+}(\zeta,\eta)|\mathbf{v}\right\rangle \langle\phi_{j,j_{z}}^{+}(\theta,\varphi)|\mathbf{r}\rangle^{*}\nonumber \\
 & \quad\times\frac{M_{j}}{M_{j}+iN_{j}}\biggr].\label{scamplitude}
\end{align}
The scattering cross section can be obtained by integrating out the
angular part of the scattered wave function,
\begin{align}
\sigma & =\int\mathrm{d}\Omega_{\bm{r}}\left|f(\mathbf{v},\bm{r})\right|^{2},\\
 & =4\pi\sum_{j}\frac{2j+1}{E^{2}}\frac{M_{j}^{2}}{M_{j}^{2}+N_{j}^{2}}.\label{eq:specificsigma}
\end{align}
For forward scattering ($\mathbf{v'=v}$), the imaginary part of the
scattering amplitude becomes 
\begin{align}
\Im f(\mathbf{v},\mathbf{v}) & =\sum_{j}\frac{2j+1}{E}\frac{M_{j}^{2}}{M_{j}^{2}+N_{j}^{2}},
\end{align}
and therefore the relation between the total scattering cross section
and the imaginary part of the forward scattering amplitude $f(\mathbf{v},\mathbf{v})$
is 
\begin{equation}
\sigma=\frac{4\pi}{E}\Im f(\mathbf{v},\mathbf{v}),\label{eq:opticaltheorem}
\end{equation}
which is the optical theorem.

\begin{figure}
\includegraphics[width=1\columnwidth]{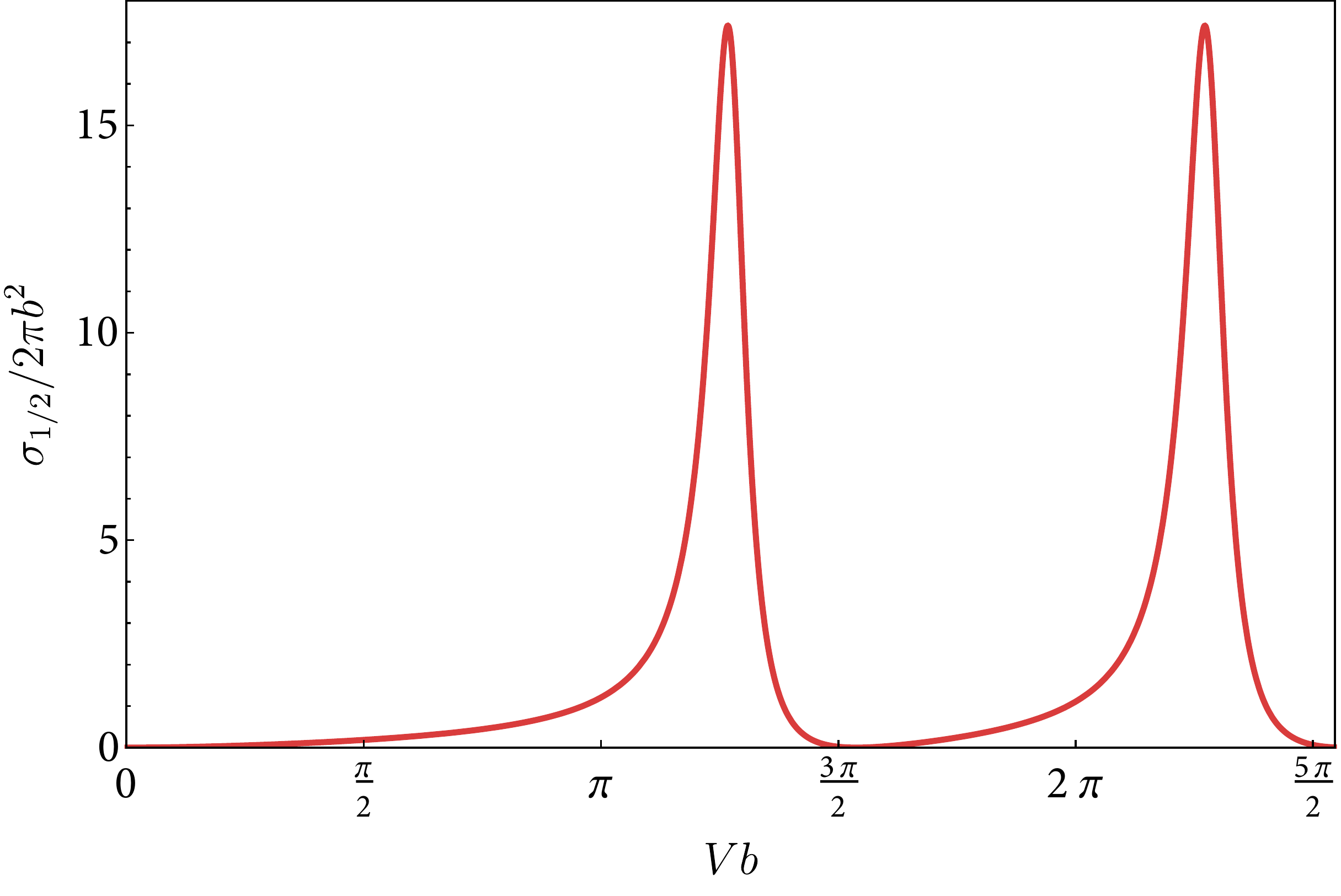}
\caption{
Total scattering cross section at low energy $E=0.5/b$ as a function
of the scaled impurity potential $Vb$, normalized to the value at $Vb=\pi$. 
At resonance the total scattering cross section peaks closely to $V_{r}b=n\pi+2Eb$, with $n$ a nonzero integer and a broadening width proportional to $(Eb)^{2}$.\label{fig:scattercrossectionvsVLow-energy-}}
\end{figure}

Fig.~\ref{fig:scattercrossectionvsVLow-energy-} shows a plot of
the total scattering cross section $\sigma_{j}$ for the lowest order
of the partial wave expansion ($j=1/2$) as a function of the impurity
strength $Vb$. 

At small $Vb$, the total scattering cross section is%
\begin{equation}
\sigma_{1/2}^{born}\sim\frac{8\pi}{9}b^{2}(Vb)^{2}(Eb)^{2},\label{eq:bornscatter}
\end{equation}
which constitutes the Born approximation. 
When $Vb$ is not small
$\sigma_{1/2}$ shows pronounced resonances whose height and  shape are independent of the order of the resonance. The peaks are located at values $V_{r}$ of the impurity potential fulfilling $V_{r}b=n\pi+2Eb+\mathcal{O}((Eb)^{2})$, where $n$ is a nonzero integer. Close to a resonance the total scattering cross section is 
\begin{equation}
\sigma_{1/2}^{res}\sim8\pi b^{2}\frac{\left(Eb\right)^{2}}{\left(Eb\right)^{4}+\left(Vb-2Eb-n\pi\right)^{2}},\label{eq:resscatter}
\end{equation}
which is obtained by first taking $Eb\rightarrow0$. 
Such a Lorentzian was also found in~\cite{Nandkishore2014}. Inserting $V_{r}$ into Eq.~(\ref{eq:resscatter}), the total scattering cross section has a maximum at $8\pi/E^{2},$ with a broadening width proportional to $(Eb)^{2}$. 
It is important to note that the resonance is not peaked at $Vb=n\pi$, which will become important in the self-consistent analysis in the next section. 
At the value $Vb=n\pi$ of the scattering strength the cross section is $\sigma_{1/2}=2\pi b^{2}$, which coincides with the value for hard wall scattering in the classical limit.

\section{Self-consistent T-matrix approach}

While the solution of the scattering problem comprises a universal resonant scattering cross section, the width of the resonance depends on the microscopic UV-cutoff $b^{-1}$. To appropriately take this into account, it is necessary to go beyond on-shell processes and include not only the energy but also the momentum dependence of the scattering amplitude. To this end, we derive in the following an explicit expression for the self-energy in the presence of finite potential scatterers within the self-consistent T-matrix approximation (SCTMA)~\cite{Ostrovsky2006}. 

\label{sec:III} 
The average Green's function of the clean system is diagonalized
in the basis of Eq.~(\ref{eq:helicitybase}). The effects of disorder are subsumed into a complex self-energy $\Sigma(E)$, which is also diagonalized in the helicity basis since the impurity is spinless. The average Green's function is therefore 
\begin{align}
G(E) & =\int\!\frac{\epsilon^2\mathrm{d}\epsilon\mathrm{d}\Omega_{\mathbf{v}}}{2\pi^3}
\frac{|\epsilon,\mathbf{v}\rangle\langle \epsilon,\mathbf{v}|}
{E-\epsilon-\Sigma(E,\epsilon)}.
\label{eq:greensfun}
\end{align}
Here, $\epsilon$
are the matrix elements of the kinetic energy in the helicity basis and $\Sigma(E,\epsilon)$ are the matrix elements of the self-energy. In the same vein, we define $G^{R/A}(E,\xi)=(E-\xi-\Sigma^{R/A}(E,\xi))^{-1}$. The eigenvalues of the bare Green's function are given by $G^{R/A}_0(E,\xi)=(E-\xi\pm i 0^+)^{-1}$.

In Sec.~\ref{sec:tmatrix} the T-matrix is computed, which allows
to calculate the self-energy for repeated scattering from a single impurity. 
By making this calculation self consistent in Sec.~\ref{sec:selfconsistent}, we can take into account certain effects from multi-impurity scattering, thus obtaining a non-vanishing imaginary part of $\Sigma$ at the
Weyl point.

\subsection{Single impurity scattering}

\begin{figure}
\includegraphics[width=\columnwidth]{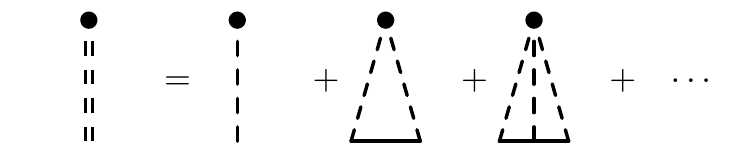}
\caption{Diagrams used in the calculation of the T-matrix. Single dashed lines are impurity scatterings from the rectangular potential, full lines are Weyl fermions and the doubly dashed line is the impurity interaction from repeated scattering. \label{fig:fullSIS}}
\end{figure}

\label{sec:tmatrix}
When the impurity concentration is low, we can treat the scattering problem of the system as that of repeated scattering off a single impurity, shown graphically in Fig.~\ref{fig:fullSIS}. 
The self-energy operator $\Sigma_0$ for single impurity scattering is by definition equal to the product of the concentration $n_{i}$ and a transition operator
$T_0$ as follows 
\begin{align}
\Sigma_0(E) & =n_{i}T_0(E),\label{eq:general_SE}
\shortintertext{which becomes obvious when writing}
T_0(E) &=\mathcal{V}+
\mathcal{V}G_{0}(E)\mathcal{V}+\dots\nonumber \\
& =\mathcal{V}+\mathcal{V}G_{0}(E)T_0(E),
\label{eq:general_Tmatrix}
\end{align}
with the bare Green's function operator $G_0(E)$. Eq.~(\ref{eq:general_Tmatrix}) can easily be solved for $T_0$ by iterating it numerically. Here, we instead opt for an analytical solution for the self-energy near to a resonance, which will guide the way to the self-consistent calculation. 

The general formalism is as follows. Given the eigenenergies $E_{\lambda}$ and eigenstates $|\psi_{\lambda}\rangle$
of the Hamiltonian Eq.~(\ref{eq:hami}), the transition operator
fulfills 
\begin{align}
\mathcal{V}|\psi_{\lambda}\rangle & =T_0(E_{\lambda})|E_{\lambda},\mathbf{v}\rangle.\label{eq:V_Tequlity}
\end{align}

Inserting the identity $\sum_{\lambda}\left|\psi_{\lambda}\right\rangle \left\langle \psi_{\lambda}\right|=1$
into Eq.~(\ref{eq:general_Tmatrix}) and making use of Eq.~(\ref{eq:V_Tequlity})
yields for $\Sigma_0(E)$~\cite{Mahan1990},
\begin{align}
\Sigma_0(E)
&=n_{i}\mathcal{V}
+n_{i}\sum_{\lambda}\frac{T_0(E_{\lambda})
\left|E_{\lambda},\mathbf{v}\right\rangle 
\left\langle E_{\lambda},\mathbf{v}\right|
T_0^{\dagger}(E_{\lambda})}
{\left(E-E_{\lambda}\right)}.\label{eq:ultimaSE}
\end{align}

We reiterate that $E$ denotes the on-shell energy, which is identical
to the energy of the incident Weyl fermion in Sec.~\ref{sec:II},
and which is given simply by the chemical potential in the WSM.
As per usual, the denominator in Eq.~\ref{eq:ultimaSE} is not well
defined unless one replaces $E\rightarrow E\pm i0^{+}$, where the signs refer to retarded and advanced self-energies, respectively.

The expectation value of the self-energy can be determined by sandwiching Eq. (\ref{eq:ultimaSE}) with a state $|\xi,\mathbf{v'}\rangle$ with scattering direction $\mathbf{v'}$ and energy $\xi$.
To this end, we write 
\begin{align}
\Sigma_0(E,\xi) & \equiv\langle \xi,\mathbf{v'}|\Sigma_0(E)|\xi,\mathbf{v'}\rangle\nonumber \\
 & =n_{i}\langle \xi,\mathbf{v'}|
 \mathcal{V}
 |\xi,\mathbf{v'}\rangle\nonumber \\
 & \qquad+n_{i}\sum_{\lambda}\frac{\left|\langle \xi,\mathbf{v'}|T_0(E_{\lambda})|E_{\lambda},\mathbf{v}\rangle\right|^{2}}{E-E_{\lambda}}.\nonumber \\
 & =n_{i}\mathcal{V}(\xi)+n_{i}\int\frac{\epsilon^{2}\mathrm{d}\epsilon\mathrm{d}\Omega_{\mathbf{v}}}{(2\pi)^{3}}\frac{\left|T_0(\xi,\epsilon,\epsilon)\right|^{2}}{E-\epsilon},\label{eq:offmassselfSEmatrixelements}
\end{align}
where we replaced $\sum_{\lambda}$ by $\int\mathrm{d}^{3}\epsilon/\left(2\pi\hbar v\right)^{3}$.
The angular integration with $\mathrm{d}\Omega_{\mathbf{v}}$ averages over all angles of incidence, which is equal to an average with respect to the position of the impurity. 
Writing out all dependencies, the T-matrix elements are defined as 
\begin{align}
T_0(\xi,E,E) & =\langle \xi,\mathbf{v'}|
T_0(E)|E,\mathbf{v}\rangle.
\end{align}
This also means that $\Sigma_0(E,\xi)=n_iT_0(\xi,E,\xi)$. For elastic scattering $(E=\xi)$, the T-matrix element can be related to the scattering amplitude as follows
\begin{align}
\langle E,\mathbf{v'}|T_0^{R}(E)|E,\mathbf{v}\rangle & =-\frac{2\pi}{E}f(\mathbf{v},\mathbf{v}'),\label{eq:defT}
\end{align}
which can be proved by use of the Lippmann-Schwinger equation.
From the imaginary part of the scattering amplitude one can therefore refer the scattering cross section.

At this point, it is necessary to assume a certain scattering potential to fill Eq.~\ref{eq:offmassselfSEmatrixelements} with life. Specifying to a rectangular impurity potential and using the spherical wave basis, we obtain immediately 
\begin{align}
\frac{\mathcal{V}(\xi) }{V} & =2\pi\sum_{j}\left(2j+1\right)\int_{0}^{b}\!\!\! r^{2}\mathrm{d}r\left[\mathscr{\textrm{j}}_{j-1/2}^{2}(\xi r)+\mathscr{\textrm{j}}_{j+1/2}^{2}(\xi r)\right]\nonumber \\
 & =2\pi b^{3}\sum_{j}\left(2j+1\right)\biggl[\mathscr{\textrm{j}}_{j-1/2}^{2}(\xi b)+\mathscr{\textrm{j}}_{j+1/2}^{2}(\xi b)\nonumber \\
 & \qquad-\frac{2j+1}{\xi b}\mathscr{\textrm{j}}_{j-1/2}(\xi b)\mathscr{\textrm{j}}_{j+1/2}(\xi b)\biggr].\label{eq:averageV}
\end{align}
From Eq.~(\ref{eq:V_Tequlity}), we can easily calculate the matrix elements
$T_0(\xi,E,E)$ with the help of the partial wave expansion of section~(\ref{sec:partialwave}).
One obtains 
\begin{align}
T_0(\xi,E,E) & =-i16\pi^{2}Vb^{3}\sum_{jj_{z}}\biggl[\frac{M_{j}(\bar{E},\xi)}{M_{j}(\bar{E},E)+iN_{j}(\bar{E},E)}\nonumber \\
 & \quad\times\frac{\langle\mathbf{v}'|j_z\rangle\langle j_z|\mathbf{v}\rangle }{(\bar{E}b-\xi b)(Eb)^{2}}\biggr],\label{eq:Tmatrixelement}
\end{align}
where we introduced the shorthand $\langle\mathbf{v}'|j_z\rangle=
\left\langle \phi_{j,j{}_{z}}^{+}(\zeta',\eta')|\mathbf{v}'\right\rangle ^{*}$, $\langle j_z|\mathbf{v}\rangle=\left\langle \phi_{j,j{}_{z}}^{+}(\zeta,\eta)|\mathbf{v}\right\rangle$ etc.
We point out that the combination $M_{j}/(M_{j}+iN_{j})$, which similarly
appeared in the scattered wave in Eq.~(\ref{scamplitude}), now contains
unequal arguments in numerator and denominator. Inserting Eq.~(\ref{eq:Tmatrixelement})
into the right hand side of Eq.~(\ref{eq:offmassselfSEmatrixelements})
and integrating out the angular dependence, we arrive at
\begin{align}
\int\!\frac{\epsilon^{2}\mathrm{d}\epsilon\mathrm{d}\Omega_{\mathbf{v}}}{\left(2\pi\right)^{3}}
\frac{\left|T_0(\xi,\epsilon,\epsilon)\right|^{2}}
{\left(E-\epsilon\right)} & =2V^{2}\sum_{j}\left(2j+1\right)I_{j}^{V}(E,\xi),\label{eq:nightmare}\\
I_{j}^{V}(E,\xi) & =\int\!\!\frac{\mathrm{d}\epsilon}{\epsilon^{2}(\bar{\epsilon}-\xi)^{2}(E-\epsilon)}\nonumber \\
 & \quad\times\frac{M_{j}^{2}(\bar{\epsilon},\xi)}{M_{j}^{2}(\bar{\epsilon},\epsilon)+N_{j}^{2}(\bar{\epsilon},\epsilon)}.\label{eq:mainintegral}
\end{align}
In the following we concentrate on the first term in the partial wave expansion, setting $j=\tfrac{1}{2}$ and henceforth dropping this index. 
Larger $j$ lead to a vanishingly small integrand in $I^{V}(E,\xi)$ near the quasiparticle pole $\epsilon=E+i0^{+}$, meaning that terms of higher angular momentum are strongly suppressed and can safely be discarded.

\subsubsection{At the limit of $Vb\rightarrow0$}

From this point on all energies occur in combination with $b$, the
size of the impurity. For the remaining analysis we therefore adopt
dimensionless variables $\tilde{E}=\frac{Eb}{\hbar v}$, $\tilde{V}=\frac{Vb}{\hbar v}$,
$\tilde{\Sigma}=\frac{\Sigma b}{\hbar v}$ etc. and $\tilde{T}=Tb^{2}\hbar v$,
$\tilde{n}_{i}=n_{i}b^{3}$, and drop the tildes. The first order
term $\langle\mathcal{V}\rangle$ in Eq.~(\ref{eq:offmassselfSEmatrixelements})
is only trivially dependent on $V$ and yields a real quantity, 
\begin{align}
\mathcal{V}(\xi) & =4\pi V\frac{\xi^{2}-\sin^{2}\xi}{\xi^{4}}.
\end{align}
For a vanishingly small $V$, the integral $I^{V}(E,\xi)$ in Eq.~(\ref{eq:nightmare}) is to leading order independent of $V$ and can be done immediately by a contour integration. 
Writing out the Bessel functions in Eq.~(\ref{eq:mainintegral}),
one obtains for $V=0$ 
\begin{align}
I^{0}(E,\xi) & =\frac{\sin^{2}\xi}{\xi^{2}}\int\!\!\frac{\mathrm{d}\epsilon}{\epsilon^{2}(\epsilon-\xi)^{2}(E-\epsilon)\xi^{2}}\bigl[\epsilon \xi\cos\epsilon\nonumber \\
 & \quad-(\xi-\epsilon+\epsilon \xi\cot \xi)\sin\epsilon\bigr]^{2}.
\end{align}
In the complex plane the integrand has poles at $\epsilon=0,E+i0^{+},\xi$.
Performing the integration of real and imaginary part of $I^{0}(E,\xi)$
yields 
\begin{align}
\Re[I^{0}] & =\frac{\pi}{2}\frac{\sin^{2}\xi}{\xi^{2}}\biggl[\frac{2}{E-\xi}\left(\frac{1}{\sin^{2}\xi}-\frac{1}{\xi^{2}}\right)\nonumber \\
 & \quad-\frac{2D\cos2E+(D^{2}-1)\sin2E}{(E-\xi)^{2}}\biggr]
 \label{eq:BA1}\\
\Im[I^{0}] & =-\pi\int\!\frac{\mathrm{d}\epsilon\delta(\epsilon-E)}{\epsilon^{2}(\epsilon-\xi)^{2}}\frac{M^{2}(\epsilon,\xi)}{M^{2}(\epsilon,\epsilon)+N^{2}(\epsilon,\epsilon)}\nonumber\\
 & =-\frac{\pi\sin^{2}\xi}{\xi^{2}}\biggl[\frac{D\sin E+\cos E}{E-\xi}\biggr]^{2},
\label{eq:BA2}
\end{align}
where $D=\xi^{-1}-E^{-1}-\cot \xi$. 
Thus, in the low energy limit by taking $E\rightarrow0$, 
\begin{align}
\Im\Sigma_0(E,\xi)=-4\pi n_{i}V^{2}E^{2}\left[\frac{\sin \xi-\xi\cos \xi}{\xi^{3}}\right]^{2}.\label{eq:SE_correction_BAlimit}
\end{align}
On the mass shell ($\xi=E$) the imaginary part of the self-energy
becomes [cf. Eq.~(\ref{eq:bornscatter})], 
\begin{align}
\Im\Sigma_0(E,E) & =-\frac{4\pi}{9}n_{i}V^{2}E^{2},\nonumber \\
\Im\frac{\Sigma_0 b}{\hbar v} & =-\frac{n_{i}b^{3}}{2}\frac{\sigma_{1/2}^{born}}{b^{2}},\label{eq:onmassshell_born}
\end{align}
where we have restored units in the last line for clarity. 
This result connects the on-mass-shell self-energy $\Sigma_0$ to the scattering amplitude $f$ and the scattering crossection $\sigma$, thus recovering the optical theorem.
We point out that Eq.~(\ref{eq:SE_correction_BAlimit}) represents $\Im\Sigma$ in the Born approximation including the momentum dependence (encoded in $\xi$). Previous calculations instead regulated the self-energy integral by introducing a cutoff scale~\cite{Hosur2012,Biswas2014}. 
The present formulation is advantageous as it allows the direct comparison with experimentally obtained parameters. 
Additionally, for weak disorder, the formulas Eqs.~(\ref{eq:BA1},\ref{eq:BA2}) hold for all $E$ and can be used to study the electronic response without further restrictions, which is however outside the present scope.

\subsubsection{Resonant scattering}

For general values of $V$, the integral in Eq. (\ref{eq:nightmare})
is 
\begin{align}
I^{V}(E,\xi) & =\frac{\sin^{2}\xi}{\xi^{2}}\int\!\frac{\mathrm{d}\epsilon}{(E-\epsilon)}\nonumber \\
 & \quad\times\frac{\left(\frac{1}{\epsilon-V}-\cot(\epsilon-V)-\frac{1}{\xi}+\cot(\xi)\right)^{2}}{(\epsilon-\xi-V)^{2}}\nonumber \\
 & \quad\times\frac{1}{1+\left(\frac{1}{\epsilon}-\frac{1}{\epsilon-V}+\cot(\epsilon-V)\right)^{2}}.
\end{align}
When the chemical potential is close to the Weyl point ($E\approx 0$), one can expand $V$ around the resonance $V_{r}$ by writing $V=n\pi+2E+\bar{\delta}$
with a small deviation $\bar{\delta}\sim E^{2}$ and a nonzero integer $n$. It is helpful to shift the integration variable $\epsilon\rightarrow\epsilon+2E+\bar{\delta}$,
leading to
\begin{align}
I^{V_{r}}(E,\xi) & =\frac{\sin^{2}\xi}{\xi^{2}}\int\!\frac{\mathrm{d}\epsilon}{(-E-\epsilon-\bar{\delta})}\nonumber \\
 & \quad\times\frac{\epsilon^{2}\left(\frac{1}{\epsilon-n\pi}-\cot(\epsilon)-\frac{1}{\xi}+\cot(\xi)\right)^{2}}{(\epsilon-\xi-n\pi)^{2}}\nonumber \\
 & \quad\times\frac{1}{\epsilon^{2}\left[1+\left(\frac{1}{\epsilon+2E+\bar{\delta}}-\frac{1}{\epsilon-n\pi}+\cot(\epsilon)\right)^{2}\right]}.\label{eq:integral_split}
\end{align}
We seek to evaluate this integral by residues. To this end, we first identify all poles in the upper half-plane of the complex variable $\epsilon$. 
In Eq.~(\ref{eq:integral_split}), the denominator in the first line defines the quasiparticle pole, located at $\epsilon=+i0^{+}-E-\bar{\delta}$.
While the second line is singular for a multiple $m$ of $\pi$, i.~e. for $\epsilon=m\pi$ (where $m\neq0,n$), these poles are canceled by zeros of the term in the third line. 
\begin{figure}
\centering\includegraphics[width=0.75\columnwidth]{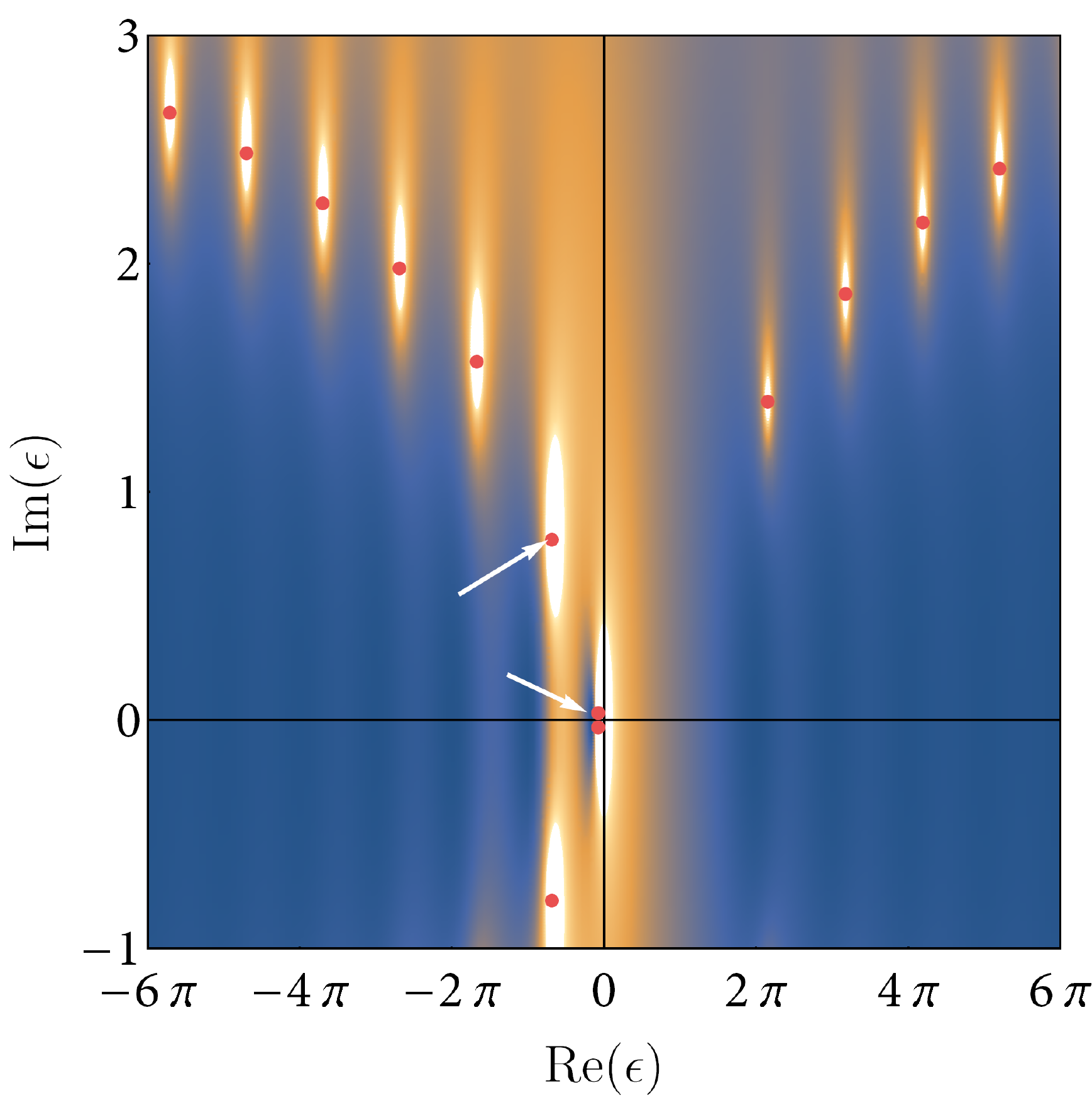}
\protect\caption{Heat map of the modulus of the third term in Eq.~(\ref{eq:integral_split})
which shows the poles in the complex plane. The red dots signify the
pole location. In the upper half plane, the two poles closest to the
origin (marked by white arrows) are the ones kept in the approximation
of Eq.~(\ref{eq:mainapproximation}). The parameters used in the
plot are $n=1$ and $2E+\bar{\delta}=0.5$.\label{fig:heatmappoles}}
\end{figure}

\begin{figure}
\includegraphics[width=1\columnwidth]{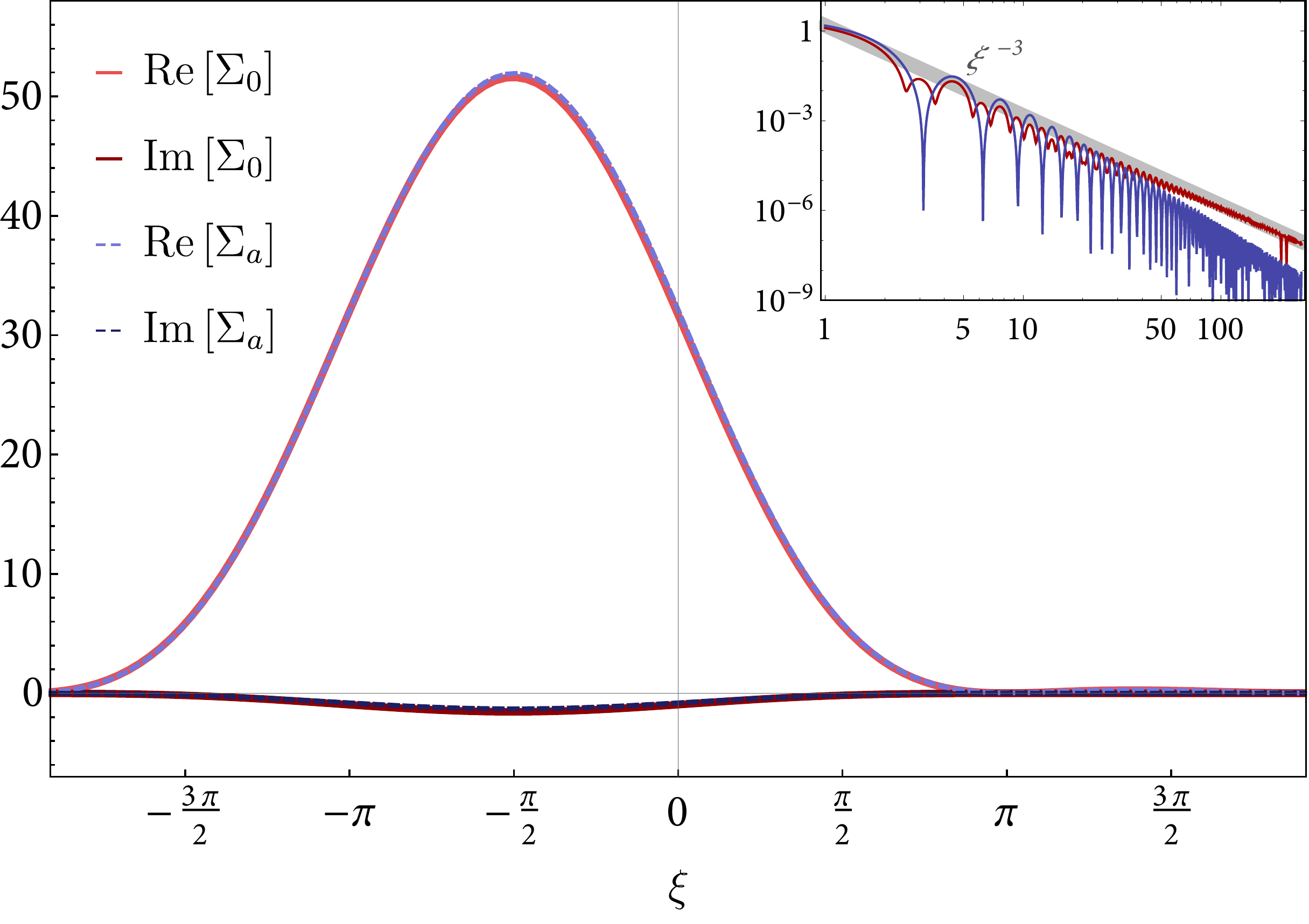} \protect\protect\caption{\label{fig:resonancezeroorderSE}Self-energy matrix elements near to resonance $V\rightarrow n\pi$ as a function of $\xi$. 
$\Sigma_0(E,\xi)$ is obtained by numerical integration of Eq.~(\ref{eq:offmassselfSEmatrixelements})
and $\Sigma_{a}$ is plotted by replacing the integral on the right
hand side of Eq. (\ref{eq:nightmare}) by the solution given in Eq.~(\ref{eq:integral_result}).
The parameters used in the plot are: $V=\pi+0.016$, $E=0.01$ and
the impurity concentration $n_{i}=0.01$. The inset shows the behavior of the modulus of the integral Eq.~(\ref{eq:nightmare}) for large $\xi$ (red) compared to the approximation of Eq.~(\ref{eq:integral_resonnance})
(blue). For better visibility parameters less close to resonance were
chosen for the inset plot.}
\end{figure}

We thus only have to take care of the poles introduced by the latter term, which are shown in Fig.~\ref{fig:heatmappoles}. 
As it turns out, the leading behavior is recovered if one keeps the pole which is closest to the origin. 
A minor complication is the fact that any approximation in the third term in Eq.~(\ref{eq:integral_split}) destroys the cancellation property between the terms of second and third line, which can be circumvented by replacing the second line
by its $\epsilon=0$ limit. 
Approximating in the third line $\cot\epsilon\approx1/\epsilon$
and $1/(\epsilon-n\pi)\approx-1/n\pi$, the poles are given by a quartic polynomial. 
Putting everything together, it is 
\begin{align}
I^{V_{r}}(E,\xi) & \approx\frac{\sin^{2}\xi}{\xi^{2}}\int\!
\frac{\mathrm{d}\epsilon}{(-E-\epsilon-\bar{\delta})}\nonumber \\
 & \quad\times\frac{1}{(\xi+n\pi)^{2}}\nonumber \\
 &\quad\times\frac{1}{\epsilon^{2}+\left(\frac{\epsilon}{\epsilon+2E+\bar{\delta}}+\frac{\epsilon}{n\pi}+1\right)^{2}}\label{eq:mainapproximation}
\end{align}
While this integral can be solved, one can further expand in the pole
closest to the origin, which yields for the third line in Eq.~(\ref{eq:mainapproximation}),
$E^{2}/(E^{4}+4(\epsilon+E+\delta/2)^{2})$, with $\delta=\bar{\delta}+E^2/n\pi$. The end result can then be written as
\begin{align}
\Sigma_0(E,\xi)&=4n_iV^2I^{V_{r}}\approx
n_i\gamma_{0}(E,\delta)t^2(\xi),
\label{eq:integral_result}
\shortintertext{where we introduced the singular amplitude $\gamma_{0}$ and a regular function $t$,}
\gamma_{0}(E,\delta) & =\frac{-4\pi}{\delta-iE^{2}}
\label{eq:integral_resonnance}\\
t(\xi)&=\frac{(n\pi)\sin\xi}{\xi(\xi+n\pi)}.
\end{align}
It is $t(0)=1$. Importantly, the resonant amplitude factorizes with respect to its energy and momentum dependence. As shown in Fig. \ref{fig:resonancezeroorderSE}, the asymptotic
solution is in a good agreement with a numerical integration. 

We can further check the validity of Eq.~(\ref{eq:integral_resonnance})
by a well defined identity on the mass-shell ($\xi=E$). To this end we use Eq.~(\ref{eq:general_SE}) to express $\Sigma_0(E,E)$ with the known matrix element $T_0(E,E,E)$ [Eq.~(\ref{eq:Tmatrixelement})],

\begin{align}
T_0(E,E,E) & =n_{i}^{-1}\Sigma_0(E,E)\nonumber \\
\frac{4\pi}{iE^{2}}\frac{M(\bar{E},E)}{M(\bar{E},E)+iN(\bar{E},E)} & =\mathcal{V}(E) +\nonumber \\
 & \hspace{-3.7cm}4V^{2}\int\!\frac{\mathrm{d}\epsilon}{\epsilon^{2}(\bar{\epsilon}-\epsilon)^{2}(E-\epsilon)}\frac{M^{2}(\bar{\epsilon},E)}{M^{2}(\bar{\epsilon},\epsilon)+N^{2}(\bar{\epsilon},\epsilon)},\label{eq:Tandsigmaequlity}
\end{align}
From this equality, we can verify both real and imaginary part of the self-energy. For the left-hand side of the identity, we obtain near to resonance 
\begin{align}
T_0(E,E,E) & =-4\pi\frac{\tan(V-2E)+iE^{2}}{\tan^{2}(V-2E)+E^{4}}.\label{eq:tmatrixreson}
\end{align}
Using the approximation of Eq.~(\ref{eq:integral_resonnance}), the
right hand side of Eq.~(\ref{eq:Tandsigmaequlity}) yields 
\begin{align}
\frac{\Sigma_0(E,E)}{n_i}
&=
\frac{4\pi}{3} V-4\pi
\frac{\delta+iE^2}{\delta^2+E^4}.
\label{eq:tmatrixreson2}
\end{align}
For a finite disorder strength ($V\approx n\pi$), the contribution $4\pi V/3$ in Eq.~(\ref{eq:tmatrixreson2}) from the impurity average $\left\langle \mathcal{V}\right\rangle $ is negligible against the contribution from resonant scattering, which diverges near to the resonance with $\delta^{-1}\sim E^{-2}$.
Keeping in mind that $V-2E\approx n\pi+\delta$, the correspondence between Eqs.~(\ref{eq:tmatrixreson}) and (\ref{eq:tmatrixreson2}) is manifest.

From the employed approximations we can deduce a restriction for $\xi$: Taking the $\epsilon\rightarrow0$ limit of the term in the second line of Eq.~(\ref{eq:integral_split}) is only valid as long as $\xi$ is not too close to a multiple of $\pi$, otherwise the $\cot \xi$ is divergent.
From the mismatch of this term with the other singular terms in the integrand we read of that this affects the integral as long as $|\xi-m\pi|<|2E+\delta|$, which can be seen in the inset plot in Fig.~\ref{fig:resonancezeroorderSE}.

We finally point out that in the limit $\xi\rightarrow\infty$ the integral $I^{V}(E,\xi)$ decays with $\xi^{-3}$, but the singular part as determined in Eq.~(\ref{eq:integral_resonnance}) actually vanishes
with $\xi^{-4}$, meaning that the effects of resonant scattering are only dominant over a finite range of energies $\xi$ (cf.~Fig.~\ref{fig:resonancezeroorderSE}). 

From the results in Eqs. (\ref{eq:onmassshell_born}) and (\ref{eq:tmatrixreson2}), we have shown that at zero chemical potential, the imaginary part of the self-energy matrix elements goes to zero, unless exactly at resonance. While one might attempt to guess the impact of resonant scattering purely on scaling arguments, tuning directly to resonance is delicate as it sensitively depends on the order of limits taken. In particular, the vanishing DOS at the Weyl point leads to an very long mean free path, so that taking only a single scattering center into account is no longer enough.
To obtain a nonzero density of states under realistic conditions it is therefore necessary to non-perturbatively include further scattering processes beyond the T-matrix approximation, which is done in the following by the use of a self-consistent calculation.

\subsection{Self-consistent T-matrix approximation}
\label{sec:selfconsistent} 
Our goal is to solve the following two equations self-consistently for $T(E)$ and $\Sigma(E)$,
\begin{align}
\Sigma(E)&=n_i T(E)\\
T(E)&=T_0(E)+T_0(E)\left[G(E)-G_0(E)\right]T(E).
\label{SCTmatrix}
\end{align}

\begin{figure}
\includegraphics[width=1\columnwidth]{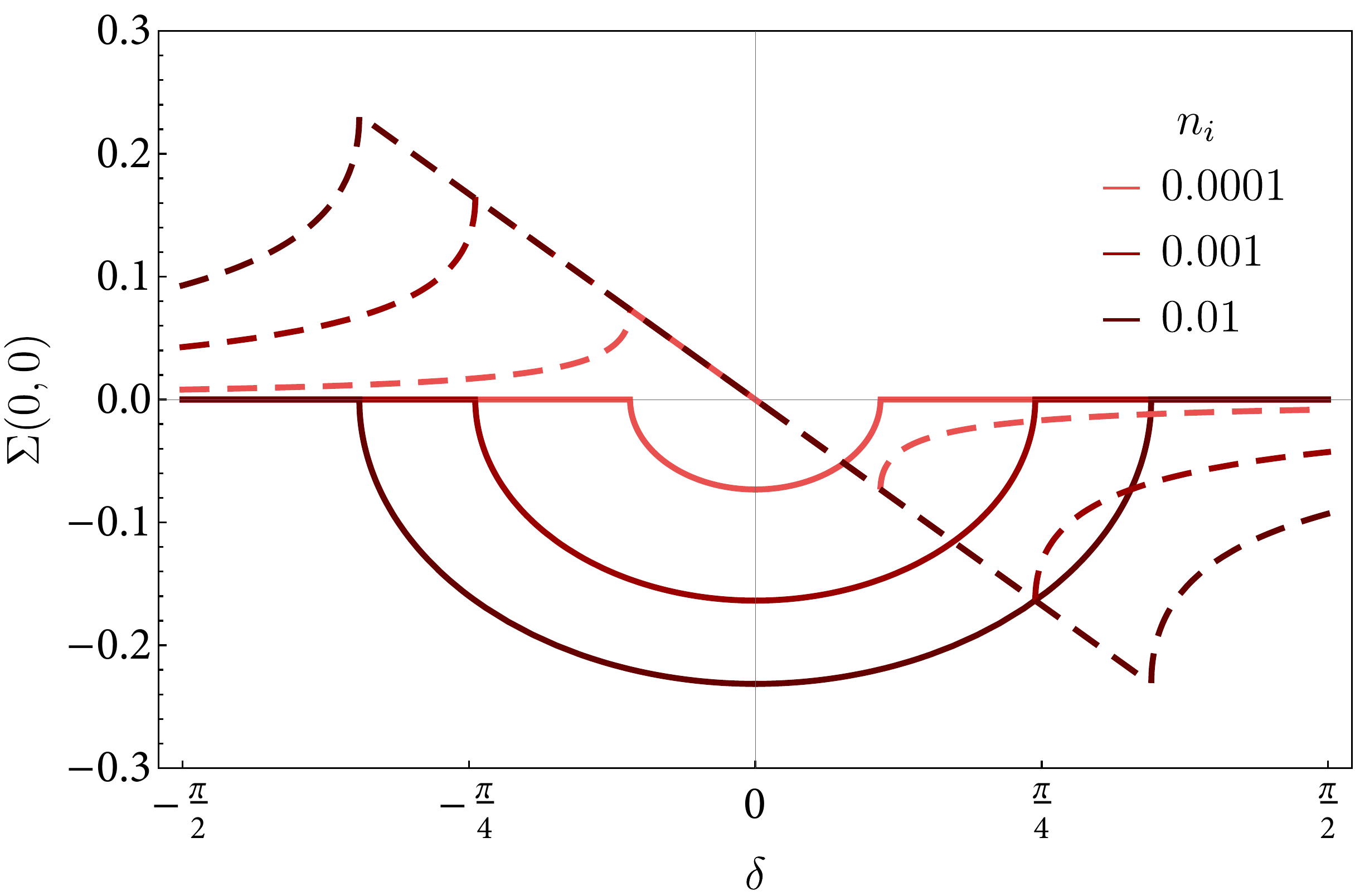} \protect\protect\caption{\label{fig:SCdetuning}
Real and imaginary part of the self-consistent self-energy (dashed and full lines, respectively) as a function of detuning. The parameters are $n=1$, $E=0$ and $\xi=0$. The approximations limit the applicability of the resonant result to values $\delta\lesssim 1$.
}
\end{figure}

Near to a resonance, the self-energy is dominated by the second term in Eq.~(\ref{eq:offmassselfSEmatrixelements}), so that it assumes the approximate form~[Eq.~(\ref{eq:integral_result})]
\begin{align}
\Sigma_0(E,\xi)&=
n_i\gamma_0(E,\delta)t^2(\xi).
\label{eq:sigma0}
\end{align}
As pointed out earlier, in this expression the dependence on $E$ and $\xi$ factorizes, which also entails that the singular part only depends on the on-shell energy $E$. This property allows for a crucial simplification of the self-consistency condition, Eq.~\ref{SCTmatrix} which we discuss now. 
The first step is to generalize the procedure used earlier in the calculation of the self-energy.
It is straightforward, but somewhat tedious to show that the  T-matrix $T_0(\xi_1,E,\xi_2)$ can indeed be calculated in the very same fashion. In the limit $E\approx 0$ and close to resonant scattering, the result is
\begin{align}
T_0(\xi_1,E,\xi_2)&=
\gamma_0(E,\delta)t(\xi_1)t(\xi_2)\nonumber\\
&\quad \times 4\pi\sum_{j_z}
\langle j_z|\mathbf{v}_2\rangle\langle \mathbf{v}_1|j_z\rangle.
\label{eq:tmatrix0}
\end{align}
Secondly, we propose for the self-consistent T-matrix $T$ and for the self-energy $\Sigma$ that the singular part $\gamma_0(E,\delta)$ in Eqs.~(\ref{eq:sigma0}), (\ref{eq:tmatrix0}) is replaced by a self-consistent amplitude $\gamma(E,\delta)$, while the momentum dependence on $\xi$ remains the same, e.~g.
\begin{align}
\Sigma(E,\xi)&=
n_i\gamma(E,\delta)t^2(\xi).
\label{eq:sigmasc}
\end{align}
Inserting this assumption into the self-consistency equation [Eq.~(\ref{SCTmatrix})] yields
\begin{align}
\gamma&=\gamma_0+4\pi\gamma_0\gamma\int \frac{\epsilon^2\mathrm{d} \epsilon }{2(2\pi)^3}
t^2(\epsilon)\left(G(E,\epsilon)-G_0(E,\epsilon)\right),
\label{eq:GG0original}
\end{align}
where the arguments of $\gamma_0$ and $\gamma$ have been suppressed. Eq.~\ref{eq:GG0original} is a much simpler integral equation which is only a function of energy.
Under the integral the combination $\epsilon^2t^2(\epsilon)$ is of order one up to a distance $n\pi$ from the origin. The separate integration of both Green's functions in the difference $G-G_0$ is therefore sensitively dependent on the UV behavior of the Green's function, and thus the precise form of $\Sigma$. Most importantly, taking $\Sigma=const.$ in the Green's function $G(E,\epsilon)$ in Eq.~(\ref{eq:GG0original}) leads to an incorrect result. This becomes obvious by forming the common denominator and using Eq.~(\ref{eq:sigmasc}), with the result
\begin{align}
\frac{\gamma}{\gamma_0}&=1+\frac{n_i \gamma^2}{4\pi^2}I_4^{0},
\label{eq:SCreduced}\\
I_4^{0}&=\int \epsilon^2\mathrm{d} \epsilon\ t^4(\epsilon)
G(E,\epsilon)G_0(E,\epsilon).
\end{align}
The integral $I_4^{0}$ contains the much faster decaying combination $\epsilon^2t^4(\epsilon)$, meaning that the self-energy in the denominator of $G$ can now be assumed as  constant. Under the assumption that $E$ and $\Sigma$ are both small, the integration then yields to leading order
\begin{align}
I_4^{0}(E=0,\Sigma=0)=\frac{ 4\pi}{3}+\frac{10}{n^2\pi}.
\label{eq:I40approx}
\end{align}
Upon approaching the resonance for a fixed impurity density $n_i$, $\gamma_0$ diverges. In this case, Eq.~(\ref{eq:SCreduced}) leads to a universal amplitude given by $\gamma^2=-4\pi^2/n_iI_4^0$. Otherwise, but as long as $I_4^{0}$ remains constant, the solution is
\begin{align}
\gamma&=2\frac{1-\sqrt{1-c_\gamma\gamma_0^2}}
{c_\gamma\gamma_0},
\qquad\quad c_\gamma=\frac{n_i I_4^0}{\pi^2}.
\label{eq:SCgeneralgamma}
\end{align}
Inserting the result near resonance into the self-consistent self-energy, one obtains a non-zero imaginary part for any finite impurity density, 
\begin{align}
\Im\Sigma(0,\xi)=-2\pi \sqrt{\frac{n_i}{I_4^0}}t^2(\xi).
\label{eq:finalscsigma}
\end{align}
The self-consistent self-energy is thus of order $\sim n_i^{1/2}$ which is indeed small, so that the approximation $\Sigma \approx 0$ which was used to evaluate Eq.~(\ref{eq:SCreduced}) is fulfilled.
The width of the resonant behavior is restricted by $1\sim c_\gamma\gamma_0^2$, which reduces to a condition for detuning $|\delta|\ll \sqrt{n_i}$ at the Weyl point. The self-energy $\Sigma(0,0)\equiv n_i\gamma$ using the function Eq.~(\ref{eq:SCgeneralgamma}) is shown in Fig.~\ref{fig:SCdetuning}.

\section{Density of states and conductivity}

\label{sec:IV} 

As it became clear in the previous section, multiple scattering leads to a finite imaginary part of the self-energy at the Weyl point. This alone implies a finite DOS and also affects the conductivity on a semiclassical level via the mean free path~\cite{Nandkishore2014}. However, with the knowledge of the momentum dependence we are now in the position to calculate these quantities based on a microscopic model. We note that the integral for the conductivity is more convergent than it was the case for the calculation of the self-energy, thus the generic form, Eq.~(\ref{eq:genericdccond}) is expected to also hold for resonant scattering. Less obvious is the expected impact of the vertex renormalization in this regime.

\subsection{Preliminaries}
Units are restored in this subsection. The density of states is given by 
\begin{align}
\rho(E) & =-\frac{1}{\pi}\Im\Tr\int\!\!\frac{\mathrm{d}^{3}k}{(2\pi\hbar)^{3}}G^{R}(E,\bm{k}).
\end{align}
Switching to the helicity basis and performing trace and angular averaging leaves 
\begin{align}
\rho(E)%
&=-\frac{1}{2(\hbar v \pi)^3}
\int\epsilon^{2}\mathrm{d}\epsilon\,\Im G^{R}(E,\epsilon).\label{eq:DOSenergies}
\end{align}
The same basis change can be applied to the Kubo formula for the longitudinal conductivity. 
Given the electron charge $e$ and the current operator
$\bm{j}=ev\bm{\sigma}$, in the limit of zero frequency and momentum the conductivity is in momentum basis
\begin{align}
\sigma_{\alpha\alpha} & =\frac{\hbar}{\pi}\Tr\int\!\!\frac{\mathrm{d}^{3}k}{(2\pi\hbar)^{3}}\left[j_{\alpha}\Im G^{R}(E,\bm{k})j_{\alpha}\Im G^{R}(E,\bm{k})\right]\nonumber \\
 & =\frac{\hbar}{4\pi}\Tr\int\!\!\frac{\mathrm{d}^{3}k}{(2\pi\hbar)^{3}}\Bigl[2j_{\alpha}G^{A}j_{\alpha}G^{R}-j_{\alpha}G^{R}j_{\alpha}G^{R}\nonumber \\
 & \qquad\qquad\qquad\qquad\quad-j_{\alpha}G^{A}j_{\alpha}G^{A}\Bigr]
 \label{eq:ARconductivity}
\end{align}

\begin{figure}
\includegraphics[width=\columnwidth]{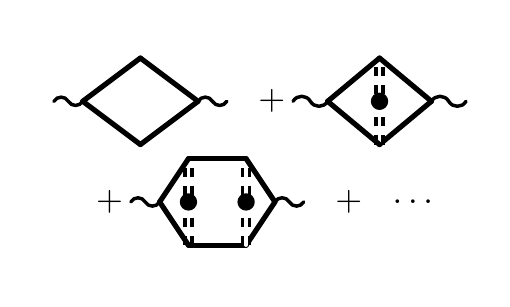}
\caption{Ladder approximation for the conductivity. The wiggly lines are attached to current vertices, the thick straight lines are dressed Weyl fermions.
\label{fig:ladder}}
\end{figure}

In the following we will employ the ladder approximation, which corresponds to the sum of diagrams shown in Fig.~\ref{fig:ladder}. 
This approximation neglects the effects of weak localization and allows to resum the series into a single bubble diagram containing a renormalized current vertex. 

Due to the momentum structure, we forgo the calculation of the renormalized vertex and instead directly compute the $N$th-order current-current correlation function $\Pi_N$ at zero frequency and momentum. Except for $N=1$, which will be treated separately, the contribution involving both advanced and retarded Green's functions is finite on its own and no UV regulation from the purely retarded or advanced diagrams is required. However, the latter parts are nonzero and cannot be neglected entirely in the quantum limit.

\begin{figure}
\centering\includegraphics[width=.7\columnwidth]{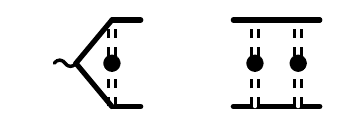}
\caption{The two building blocks of the conductivity calculation. The symbols are the same as in Fig.~\ref{fig:ladder}.
\label{fig:PiN}}
\end{figure}

The units for DOS and conductivity are $[\rho]=(\hbar v b^2)^{-1}$ and
$[\sigma_{\alpha\alpha}]=e^{2}/ \hbar b$.

\subsection{Density of states at the Weyl point}

\begin{figure}
\includegraphics[width=\columnwidth]{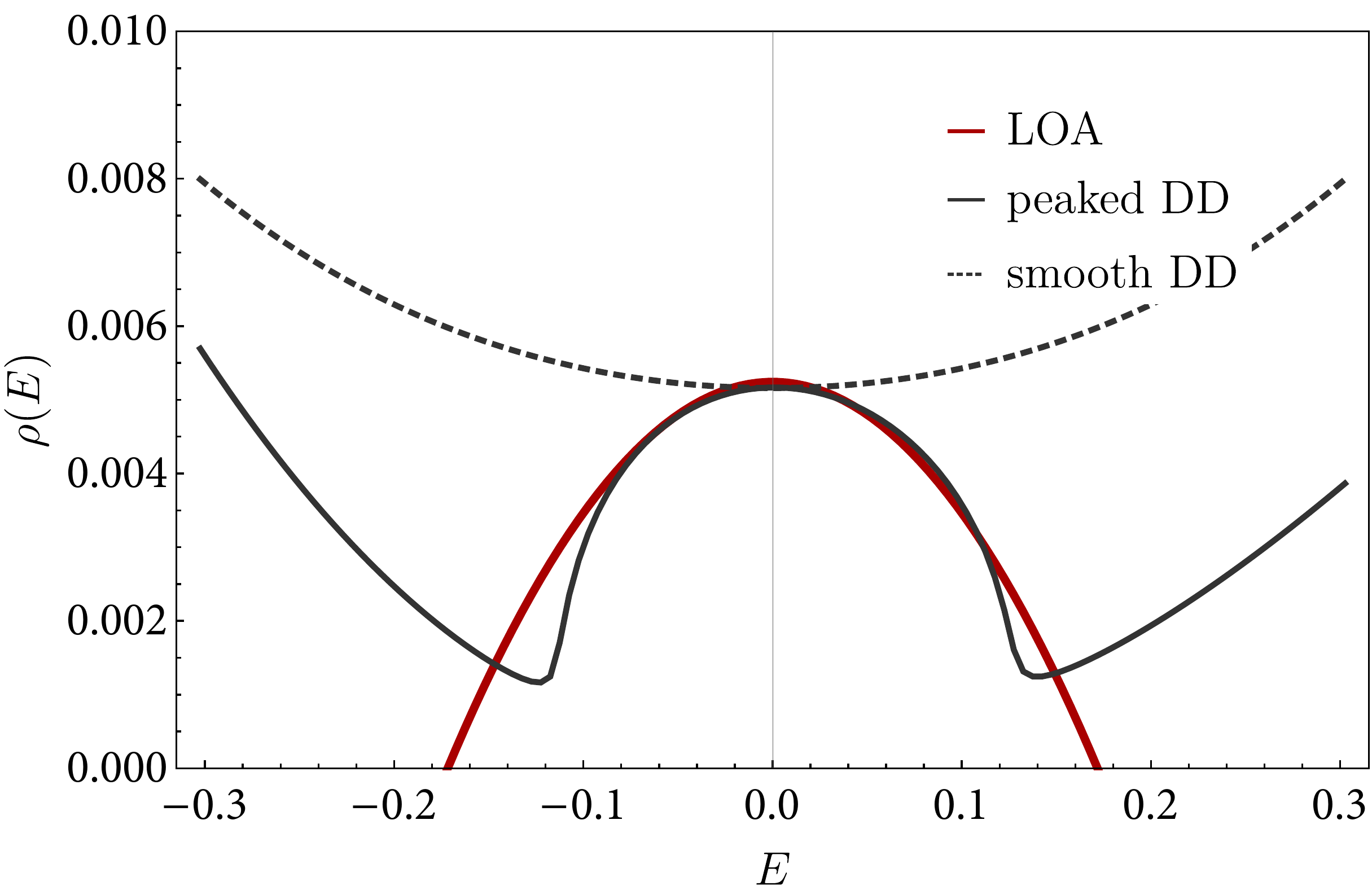}
\caption{DOS as a function of chemical potential, where $E=0$ is the Weyl point. The red line indicates the DOS induced by resonant scattering from rare regions of strength $V=\pi$ in the leading order approximation (LOA), Eq.~\ref{eq:DOSLOA}. Plotting Eq.~\ref{eq:DOSresult1} directly yields the gray curve, which is not particle hole symmetric, as the disorder distribution (DD) contains only impurities with the same potential. Here, the resonance is centered around $E=0$ for a choice $V=\pi+0.026$, the deviation from $\pi$ being due to a small shift in the resonance condition for the self-consistent case. A smooth DD would yield something close to the dashed line, where the resonance peak is smeared out. Parameters are $n_i=5\cdot 10^{-4}$, $n=1$.}
\label{fig:DOSbump}
\end{figure}

It holds generally that $\Im G^{R}=\Im\Sigma^RG^RG^A$.
For nearly resonant scattering the self-energy is dominated by its imaginary part, since the real part vanishes at resonance [Eq.~(\ref{eq:finalscsigma})]. Thus, we proceed to drop the real part of the self-energy and estimate the DOS as
\begin{align}
\rho(E) & =-\frac{1}{2\pi^{3}}\int\!\epsilon^{2}\mathrm{d}\epsilon\,
\frac{\Im\Sigma^R(E,\epsilon)}{(E-\epsilon)^{2}+[\Im\Sigma^R(E,\epsilon)]^{2}}.
\end{align}
Within the SCTMA approximation for resonant scattering which was detailed in Sec~\ref{sec:III}, the self-energy factorizes with respect to energy and momentum. Consequently, in the integration the dependence on $\epsilon$ of the self-energy is regular, leading to a vanishing integrand near to $\epsilon\approx0$. Thus, the integral is \emph{not} dominated by the contribution of the quasiparticle pole, but by the behavior around $\epsilon=-n\pi$.
Approximating the self-energy in the numerator by its resonant form, $\Im\Sigma^R(E,\epsilon)=n_i\Im\gamma(E,\delta) t^2(\epsilon)$ and in the denominator by its value at $\epsilon=-n\pi$, 
\begin{align}
-\Im\Sigma^R(-n\pi,0)=-n_i\Im\gamma(E,\delta)\equiv i s, 
\end{align}
which also coincides with the value of $\Sigma$ at $\epsilon=0$, we obtain
\begin{align}
\rho(E)&=
\frac{s}{2\pi^3}I_2(E,s),
\label{eq:DOSresult1}\\
I_2(E,s)&=
\int\!\epsilon^{2}\mathrm{d}\epsilon\, t^2(\epsilon)G^R_s(E,\epsilon)G^A_s(E,\epsilon).
\label{eq:defI2}
\end{align}
In this expression, we introduced a simplified form of the Green's function, $G_s^{R/A}(E,\xi)=(E-\xi\pm i s)$. 
It is important to note that similar to  Eq.~(\ref{eq:GG0original}), convergence is provided by the momentum dependence of the self-energy in the numerator. 
The evaluation of $I_2$ is written out in appendix~\ref{sec:energyints}.
At the Weyl point and for vanishing impurity density, it is $I_2(0,0)=2\pi$, the density of states is thus of size $\rho(0)\sim \sqrt{n_i}$.
We point out that the condition for resonant scattering was $V_r=n\pi+2E+\mathcal{O}(E^2)$.
Therefore, the leading effect of a change in the chemical potential will be a detuning away from resonance. In particular, if the WSM contains impurities with a given potential which becomes resonant for $E=0$, the DOS develops a characteristic bump,
\begin{align}
\rho(E)&=-\frac{n_i}{\pi^2}\Im \gamma(E,0)\\
&\approx c_\rho \sqrt{n_i}-\frac{c_\rho'E^2}{\sqrt{n_i}}.
\label{eq:DOSLOA}
\end{align}
For an impurity potential $V=\pi$, it is $c_\rho=2(I_4^0)^{-1/2}/\pi=0.23$ and $c_\rho'=(I_4^0)^{-3/2}/4\pi=0.004$. At any finite disorder density $n_i$, there is a small additional shift of the resonance condition in the self-consistent calculation due to the change of $I_4^0$ with nonzero self-energy (cf. appendix ~\ref{sec:energyints}).
In many cases, the disorder is instead assumed to possess a smooth distribution function in the impurity potential strength. Then the bump in the DOS is smeared out, leaving only a bottoming-out of the DOS at a finite value. This was previously seen in numerical calculations using Gaussian or box disorder~\cite{Pixley2016a}.
The DOS in the presence of rare regions is shown in Fig.~\ref{fig:DOSbump}. 

\subsection{Disorder enhanced conductivity}

While the Green's function is diagonal in the helicity basis, both the T-matrix $T$ and the current operator $j_\alpha$ are not. 
Thus, each fermion line in the current bubble can be of  different helicity. 
Additionally, each T-matrix introduces a sum over angular indices.
In the following, we compute the conductivity via $\tfrac{1}{3}(\sigma_{xx}+\sigma_{yy}+\sigma_{zz})$ in order to simplify some intermediate steps. 
For the self-consistent self-energy at resonance the approximate expression of Eq.~(\ref{eq:finalscsigma}) will be used. 

The T-matrix approximately factorizes with respect to both, the angular and the energy dependence of the in- and outgoing states. 
This makes it possible to calculate each loop appearing in $\Pi_N$ independently, leaving only two distinct building blocks (Fig.~\ref{fig:PiN}), which  resemble a one-loop current vertex  and a four-point vertex, respectively.

Setting $j=1/2$, for $N=1$, it is necessary to keep all three contributions in Eq.~(\ref{eq:ARconductivity}) from the start to ensure convergence. The current vertices can be resolved straightforwardly:
\begin{align}
&\langle \mathbf{v}|\bm{\sigma}|\pm\mathbf{v}\rangle
\langle \pm\mathbf{v}|\bm{\sigma}|\mathbf{v}\rangle
\nonumber\\&=
\Tr\left[\bm{\sigma}|\pm\mathbf{v}\rangle
\langle \pm\mathbf{v}|\bm{\sigma}|\mathbf{v}\rangle
\langle \mathbf{v}|\right]
\nonumber\\&=
\tfrac{1}{4}\Tr\left[
\bm{\sigma}\cdot \bm{\sigma}\pm
\bm{\sigma}(\bm{\sigma}\cdot\mathbf{v})
\bm{\sigma}(\bm{\sigma}\cdot\mathbf{v})
\right]
\nonumber\\&=\tfrac{1}{4}(6\mp2|\mathbf{v}|^2)=\frac{3\mp 1}{2}.
\end{align}
After the angular integration one thus arrives at 
\begin{align}
\Pi_1 & =\frac{1}{6\pi^{2}}\int\!\epsilon^{2}\mathrm{d}\epsilon\Bigl[\left(\Im G^{R}(E,\epsilon)\right)^{2}\nonumber \\
 & \quad+2\Im G^{R}(E,\epsilon)\Im G^{R}(E,-\epsilon)\Bigr],
\end{align}
where retarded and advanced functions have already been recombined. Pulling out the energy integrals, this becomes
\begin{align}
\Pi_1 & =\frac{s^2}{6\pi^{2}}
(I_4''+2I_4''')\\
I_4''&=\int\epsilon^{2}\mathrm{d}\epsilon\,
t(\epsilon)^4[G^R_s(E,\epsilon)]^2[G^A_s(E,\epsilon)]^2\\
I_4'''&=\int\epsilon^{2}\mathrm{d}\epsilon\,
\bigl[t^2(\epsilon)G^R_s(E,\epsilon)G^A_s(E,\epsilon)
\nonumber\\&\quad\times
t^2(-\epsilon)G^R_s(E,-\epsilon)G^A_s(E,-\epsilon)\bigr].
\end{align}
Close to the Weyl point and to leading order in the impurity density $n_i$, we obtain 
\begin{align}
I_4''&=\frac{\pi}{2s}+\frac{\pi E^2}{2s^3}\\
I_4'''&=\frac{\pi}{2s},
\end{align}
which results in 
\begin{align}
\sigma_1&=\frac{1}{12\pi^2}\frac{E^2+3s^2}{s}
\end{align}
This reproduces the result recently obtained for a generic energy dependent impurity scattering rate model, Eq. (\ref{eq:genericdccond}), including the numerical coefficient~\cite{Tabert2016}. We thus conclude that even for resonant scattering, the conductivity as given by the Kubo formula does not depend on the microscopic details of a short-ranged impurity except through the self-energy $\Sigma(0,0)$. 

The ladder term $\widetilde{P}$ with external legs with indices $1,2,3,4$ and zero momentum transfer ($\mathbf{v}_1\xi_1=\mathbf{v}_4 \xi_4$, $\mathbf{v}_2\xi_2=\mathbf{v}_3 \xi_3$) is given by
\begin{widetext}
\begin{align}
\widetilde{P}(\xi_1,\xi_2,\xi_3,\xi_4)
&=n_i^2\langle\xi_1,\mathbf{v}_1|
\langle\xi_4,\mathbf{v}_4|
T(E)G^R(E)T^\dagger(E)T(E)G^A(E)T^\dagger(E)
|\xi_2,\mathbf{v}_2\rangle
|\xi_3,\mathbf{v}_3\rangle
\\
&=n_i|4\pi\gamma|^2
t(\xi_1)t(\xi_2)t(\xi_3)t(\xi_4)
\smashoperator{\sum_{j_{z1},j_{z2},j_{z3},j_{z4}}}{}
\langle\mathbf{v}_1|{j_{z1}}\rangle
\langle {j_{z2}}|\mathbf{v}_2\rangle
\langle\mathbf{v}_3| {j_{z3}}\rangle
\langle {j_{z4}}|\mathbf{v}_4\rangle
\hat{P}_{j_{z1}j_{z2}j_{z3}j_{z4}}
\notag\\
\hat{P}_{j_{z1}j_{z2}j_{z3}j_{z4}}&=
n_i|4\pi\gamma|^2
\int\!\! \frac{\epsilon^2\mathrm{d}\epsilon}{(2\pi)^3}
t(\epsilon)^4G^R(E,\epsilon)G^A(E,\epsilon)
\int \mathrm{d}\Omega_{\mathbf{v}} \langle {j_{z1}}|\mathbf{v}\rangle
\langle\mathbf{v}| {j_{z2}}\rangle
\langle {j_{z3}}|\mathbf{v}\rangle
\langle\mathbf{v}| {j_{z4}}\rangle+\dots
\end{align}
\end{widetext}
Here, the dots indicate that $\epsilon$ and $\mathbf{v}$ can be of dissimilar helicity in the upper and the lower line. 
The angular integration yields 
\begin{align}
\hat{P}_{1234}&=
(P+2P')\delta_{1,2}\delta_{3,4}
\notag\\&\quad
+(P-P')\delta_{1,4}\delta_{23},
\label{eq:ladderelem}
\end{align}
with
\begin{align}
P&=\frac{n_i|\gamma|^2}{12\pi^2}I_4(E,s)\qquad
P'=\frac{n_i|\gamma|^2}{12\pi^2}I_4'(E,s),
\end{align}
\begin{align}
I_4(E,s)&=\int\! \epsilon^2\mathrm{d}\epsilon\,
t(\epsilon)^4G^R_s(E,\epsilon)G^A_s(E,\epsilon)
\\
I_4'(E,s)&=\int\! \epsilon^2\mathrm{d}\epsilon\,
t(\epsilon)^2t(-\epsilon)^2
G^R_s(E,-\epsilon)G^A_s(E,\epsilon).
\end{align}
Based on the structure of $\hat{P}$ we propose for the ladder sum a kernel of the form
\begin{align}
D\delta_{1,2}\delta_{3,4}
+D'\delta_{1,4}\delta_{2,3}.
\end{align}
The recursion relation for the ladder sum then reads
\begin{align}
& D\delta_{1,2}\delta_{3,4}+
D'\delta_{1,4}\delta_{2,3}
\notag\\&=
\delta_{1,1}\delta_{3,4}
+
\sum_{a,b}
\Bigl[(D\delta_{a,2}\delta_{3,b}+
D'\delta_{a,b}\delta_{2,3})
\notag\\&\qquad\qquad\qquad\qquad\times 
\bigl((P+2P')\delta_{1,a}\delta_{b,4}
\notag\\&\qquad\qquad\qquad\qquad+
(P-P')\delta_{1,4}
\delta_{a,b}\bigr)\Bigr],
\end{align}
the solution of which is
\begin{align}
D&=\frac{1}{1-P-2P'}\\
D'&=\frac{P-P'}{(1-3P)(1-P-2P')}.
\end{align}
The ladder is closed with a current vertex on both ends, with the closing loop containing the following integrals,
\begin{align}
\bm{C}_{j_{z1}j_{z4}}&=\biggl[\int \frac{\epsilon^2\mathrm{d}\epsilon}{(2\pi)^3}
[t(\epsilon)^2G^R(E,\epsilon)G^A(E,\epsilon)
\notag\\&\quad
\int \mathrm{d}\Omega_{\mathbf{v}}
\langle {j_{z1}}|\mathbf{v}\rangle 
\langle \mathbf{v}|\bm{\sigma}|\mathbf{v}\rangle
\langle \mathbf{v}| {j_{z4}}\rangle+\dots\biggr],
\end{align}
where the dots again denote the summation over different helicities. The scalar product of left and right termination is
\begin{align}
\bm{C}_{14}\cdot
\bm{C}_{23}
&=\left(2\delta_{1,2}\delta_{3,4}
-\delta_{1,4}\delta_{2,3}\right)
\notag\\&\quad\times
\frac{1}{(2\pi)^6}\left(\frac{I_2}{6}+\frac{I_2'}{3}\right)^2.
\end{align}
$I_2$ was already defined in~(\ref{eq:defI2}), $I_2'$ is
\begin{align}
I_2'(E,s)&=\int\!\epsilon^2\mathrm{d}\epsilon\,
t(\epsilon)t(-\epsilon)
G^R_s(E,-\epsilon)G^A_s(E,\epsilon).
\end{align}
The energy integrals $I_4$, $I_4'$, $I_2$ and $I_2'$ are tabulated in appendix~\ref{sec:energyints}. For $E\approx0$ and $s\sim E$ they reduce to
\begin{align}
I_4(E,s)&=\frac{4\pi}{3}+\frac{10}{n^2\pi}\label{eq:diffpole}\\
I_4'(E,s)&=-\frac{2\pi}{3}-\frac{5}{4n^2\pi}\\
I_2(E,s)&=2\pi-\pi s-\frac{6E}{n}+\frac{\pi E^2}{s}\\
I_2'(E,s)&=-\pi+2\pi s
\end{align}
The ladder sum is then
\begin{align}
3\Pi^{AR}&=\frac{n_i|4\pi\gamma|^2}{6(2\pi)^6}\frac{(I_2+2I_2')^2}{1-P-2P'}\\
\Pi^{AR}&\approx\frac{n_i|\gamma|^2}{72\pi^4}\frac{(3\pi s+\pi E^2/s-6E/n)^2}{1-c_I},
\label{eq:laddersum}
\end{align}
with $c_I=5n_i|\gamma|^2/8n^2\pi^3$. 
It was shown in Sec.~\ref{sec:selfconsistent} that the amplitude squared of the T-matrix approaches  $\gamma^2=-4\pi^2/n_iI_4^0$ at resonance , which means that for $n=1$ it is $c_I=0.108$. As a consistency check we show that these expressions reproduce the diffusion pole, confirming that the approximations taken for the determination of the T-matrix conserve charge.
For the diffuson, the terminating angular integrations yield a term $\sim\delta_{1,4}\delta_{2,3}$. After performing the trace, the result is thus proportional to
\begin{align}
\frac{1}{1-3P}.
\end{align}
The diffusion pole is recovered for $P=1/3$, which is true if $I_4=I_4^0$, as it is indeed the case to leading order in $E$ and $s$ [Eqs.~(\ref{eq:I40approx}, \ref{eq:diffpole})].
It is also possible to show more generally that the corresponding Ward identity is conserved asymptotically for $E\approx 0$, which is detailed in appendix~\ref{app:ward}.

We also calculated the doubly retarded ladder, which is of comparable size at the Weyl point. The result is
\begin{align}
\Pi^{RR}&=\frac{n_i|\gamma|^2}{72\pi^4}\frac{\left(\frac{6(E+is)}{n}\right)^2}{1-c_I}.
\end{align}

As it becomes clear from Eq.~(\ref{eq:laddersum}), the ladder sum is of order $s^2$ and thus subleading compared to the particle-hole bubble $\Pi_1$ if one approaches the Weyl point, $E\rightarrow 0$. This surprising behavior is not linked to the ladder being perturbatively irrelevant, at resonance each element is indeed of order 1 which results in a finite renormalization by $1-c_I$ in Eq.~(\ref{eq:laddersum}). Instead, a factor proportional to $s$ is introduced by each of the two enclosing current vertices. 
This leads to the rather unusual situation that the semimetallicity becomes partially masked close to the Weyl point. In particular, the conductivity is given just by the Kubo formula, like in a metal with isotropic scattering. On the other hand, the diffusion coefficient is depending on $I_4$, where the special properties of the scattering matrix for a Weyl dispersion enters explicitly.

In summary, using a self-consistent self-energy, the vertex renormalization has no effect to leading order in the impurity density, in spite of the fact that the scattering matrix  remains anisotropic. 
This finding applies exclusively to the quantum limit, in the limit where the Boltzmann approach works and $E^2/s$ is much larger than $s$, the scattering amplitude is $\gamma_0$ and the ladder leads to the expected replacement of the quasiparticle lifetime $\tau$ by transport lifetime $\tau_{tr}$ in the expression for the conductivity. For resonant scattering and if $E^2 \gg s$, the ratio is
\begin{align}
\frac{\tau_{tr}}{\tau}&=1+\frac{1}{3(1-\frac{n_i|\gamma_0|^2}{12\pi^2}\frac{\pi E^2}{s})}=\frac{3}{2}.
\end{align}
We note that this result has only a limited region of applicability, as resonant scattering eventually becomes negligible for a large enough DOS away from the Weyl point. In particular, there is not necessarily any intermediate regime where the scattering rate due to a T-matrix with amplitude $\gamma_0$ is larger than the one given by the SCBA. Importantly, the same ratio $\tau_{tr}/\tau=3/2$ was earlier obtained within SCBA~\cite{Biswas2014}.

Putting everything together and restoring units, in the presence of resonant scattering the conductivity is at zero frequency given by
\begin{align}
\sigma_{\alpha\alpha}
&=\frac{e^2}{\hbar^2 v}\frac{3s^2+E^2}{12\pi^2 s},
\label{eq:finalconductivity}
\end{align}
where the scattering rate in units of energy is $s=2\pi\hbar v\sqrt{n_ib/I_4^0}$.
This result was obtained using a rectangular impurity potential, but we do not expect that a generic short-ranged potential will change the expression. 
This can be concluded from the value of $I_4^0$, which is essentially determined by the large scattering amplitudes at $E=0$ and $Eb=-n\pi\hbar v$ and does in leading order not depend on the microscopic details of the impurity.
However, we point out that the conductivity is not dimensionless in 3D, therefore the length scale $b$ still enters explicitly in Eq.~(\ref{eq:finalconductivity}). For a general impurity potential this length scale is given by the effective size of an impurity site. 

Compared to the previous estimate of the mean free path for hopping between rare regions $\ell\sim (n_i b^2)^{-1}$~\cite{Nandkishore2014}, it turned out to be of size $(n_ib)^{-1/2}$. In the presence of disorder with an Gaussian impurity potential distribution this finding implies that the crossover from the SCBA regime to one dominated by resonant scattering happens at comparatively higher scales than previously thought. 
As one important consequence we estimate how the conductivity saturates near the Weyl point as a function of decreasing chemical potential. As discussed in the calculation of the DOS, a smooth disorder distribution smears out the effects of detuning. The mean free path due to resonant scattering is shorter than the one given in SCBA once $E<\hbar v(n_ib)^{1/4}$. Continuing to smaller scales, Eq.~\ref{eq:finalconductivity} tells us that the semiclassical term $E^2/s$ is comparable to the part of size $s$ if $s\sim E$, i.e. it is irrelevant once $E<\hbar v\sqrt{n_ib}$. Below this scale, the conductivity approaches a residual value ($n=1$)
\begin{align}
\sigma_{min}&=0.06 N\sqrt{n_i b}\frac{e^2}{\hbar} .
\label{eq:rescond}
\end{align}
So far the focus was on a single Weyl cone ($N=1$), in any real material $N$ is even and can be as large as 24~\cite{Wan2011}.

\section{Conclusion}
\label{sec:V}
We have presented the solution to the scattering problem of chiral fermions in a WSM close to resonant scattering. 
The singular amplitude of the T-matrix was found to factorize with respect to energy and momentum dependence, which made it possible to also determine the T-matrix in a self-consistent fashion. Using the self-consistent solution, which is non-analytic in disorder strength, a finite DOS emerges at the Weyl point. Interestingly, the lifting of the semimetallic DOS turns out to be large enough to render vertex corrections to the Kubo estimate of the conductivity to be irrelevant. This was demonstrated in a conserving approximation. 

The main consequence for experiment is the predicted residual conductivity at charge neutrality [Eq.~(\ref{eq:rescond})], which scales with the square root of the impurity density and is thus qualitatively larger than previously estimated. 

Apart from the consequences for zero temperature presented here, in the future the same calculational scheme can be easily adapted to calculate finite T and finite-$\omega$ transport properties in a microscopic model. This can complement studies with generic scattering models~\cite{Lundgren2014,Tabert2016} by supplying a detailed understanding of the underlying scattering amplitudes. Another interesting possibility is the study of boundary effects, which is important to estimate the stability of topological properties of a dirty WSM.

\section*{Acknowledgments}
We thank E. Khalaf for useful discussions. T. H. is supported by the Minerva foundation.

\appendix

\section{Partial wave expansion}

\label{sec:appA} The scattering theory for particles with spin $1/2$ is well-known, see e.~g.~\cite{Sakurai2011}. For completeness we note here the main ingredients used to derive Eqs.~(\ref{eq:totalwavefunction}), (\ref{scamplitude}) and (\ref{eq:specificsigma}) of the main text.
Scattering is generally treated by taking an incident plane wave $|E,\mathbf{v}\rangle$ as the initial state, expanding this wave in a spherical basis $|\phi_{j,j_{z}}\rangle$ to solve the scattering problem itself, and then constructing an outgoing plane wave $|E,\mathbf{v'}\rangle$ for the final state. Using dimensionless variables and writing the Hamilton operator as 
\begin{align}
H & =v\bm{\sigma}\cdot\bm{p}+\mathcal{V}(r)\nonumber \\
 & =-i\hbar v\bm{\sigma}\cdot\hat{\bm{r}}\left(\frac{\partial}{\partial r}-\bm{\sigma}\cdot\bm{L}\right)+\mathcal{V}(r),\label{eq:hami2}
\end{align}
where $\bm{L}$ is the orbital angular momentum, it becomes clear
that $H$ conserves the total angular momentum $\bm{J}=\bm{L}+\bm{\sigma}/2$ but not $\bm{L}$. 
To construct the eigenstates, we proceed with a decomposition into a radial part $R$ and an angular part $|\phi\rangle$ which is a two component spinor, $|\psi\rangle=R(r)|\phi(\theta,\varphi)\rangle$.
If a Hamiltonian preserves separately spin, angular momentum and total angular momentum, the eigenbasis of the angular part is given by the spherical spinors $|\phi_{j,j_{z}}^{+}\rangle$ and $|\phi_{j,j_{z}}^{-}\rangle$ which are defined as~\cite{Sakurai2011} \begin{align}
|\phi_{j,j_{z}}^{+}\left(\theta,\varphi\right)\rangle & =\left(\begin{array}{c}
\sqrt{\frac{j+j_{z}}{2j}}\mathsf{\mathbb{\mathcal{\mathscr{Y}}}}_{j_{z}-1/2}^{j-1/2}(\theta,\varphi)\\
\sqrt{\frac{j-j_{z}}{2j}}\mathscr{Y}_{j_{z}+1/2}^{j-1/2}(\theta,\varphi)
\end{array}\right),\shortintertext{and}|\phi_{j,j_{z}}^{-}\left(\theta,\varphi\right)\rangle & =\left(\begin{array}{c}
-\sqrt{\frac{j-j_{z}+1}{2j+2}}\mathscr{Y}_{j_{z}-1/2}^{j+1/2}(\theta,\varphi)\\
\sqrt{\frac{j+j_{z}+1}{2j+2}}\mathscr{Y}_{j_{z}+1/2}^{j+1/2}(\theta,\varphi)
\end{array}\right).\label{eq:phim2}
\end{align}
The convention used here is $\mathscr{Y}_{m}^{l}(\theta,\varphi)=\sqrt{\left(l-m\right)!\left(2l+1\right)/4\pi\left(l+m\right)!}\, e^{im\varphi}P_{l}^{m}\left(\cos\theta\right)$
for the spherical harmonic function, and $P_{l}^{m}$ is the associated
Legendre function. The basis functions $|\phi_{j,j_{z}}^{\pm}\rangle$
satisfy the relations
\begin{align}
\bm{\sigma}\cdot\bm{L}|\phi_{j,j_{z}}^{+}\rangle
&=\hbar\left(j-1/2\right)|\phi_{j,j_{z}}^{+}\rangle,\\
\bm{\sigma}\cdot\bm{L}|\phi_{j,j_{z}}^{-}\rangle
&=-\hbar(j+3/2)|\phi_{j,j_{z}}^{-}\rangle,\\
\bm{\sigma}\cdot\hat{\bm{r}}|\phi_{j,j_{z}}^{+}\rangle
&=-|\phi_{j,j_{z}}^{-}\rangle,\\
\bm{\sigma}\cdot\hat{\bm{r}}|\phi_{j,j_{z}}^{-}\rangle
&=-|\phi_{j,j_{z}}^{+}\rangle.
\end{align}
The clean Weyl Hamiltonian can thus be solved with the ansatz 
\begin{align}
|E,j,j_{z}\rangle & =f(r)|\phi_{j,j_{z}}^{+}\rangle-ig(r)|\phi_{j,j_{z}}^{-}\rangle,
\end{align}
yielding two coupled differential equations for the radial part, 
\begin{eqnarray}
\frac{\partial f(r)}{\partial r}-\left(j-\frac{1}{2}\right)\frac{f(r)}{r} & = & -Eg(r),\allowdisplaybreaks[1]\label{eq:gr2}\\
\frac{\partial g(r)}{\partial r}+(j+\frac{3}{2})\frac{g(r)}{r} & = & Ef(r).\label{eq:fr2}
\end{eqnarray}
The solutions are 
\begin{align}
f(r) & =A_{j}\mathscr{\textrm{j}}_{j-1/2}(Er)+B_{j}\mathscr{\textrm{n}}_{j-1/2}(Er),\label{eq:f2}
\shortintertext{and}
g(r) & =A_{j}\mathscr{\textrm{j}}_{j+1/2}(Er)+B_{j}\mathscr{\textrm{n}}_{j+1/2}(Er),\label{eq:g2}
\end{align}
where $\mathscr{\textrm{j}}_{l}(x)$ and $\mathscr{\textrm{n}}_{l}(x)$
are the spherical Bessel functions of the first and the second kind.
For a rectangular impurity potential $\mathcal{V}(r)$, Eqs.~(\ref{eq:f2})
and (\ref{eq:g2}), which are valid for any flat potential, can be
immediately used to construct the wave function inside and outside
of the impurity, leading to Eq.~(\ref{eq:totalwavefunction}).

We now turn to the scattering problem. The total wave function in
the presence of an impurity was already defined in Eq.~(\ref{eq:scatteringwf})
in the main text: 
\begin{align}
|\psi\rangle & =e^{iE\mathbf{v}\cdot\bm{r}}|\mathbf{v}\rangle+f(\mathbf{v},\hat{\bm{r}})\frac{e^{iEr}}{r}|\hat{\bm{r}}\rangle.\label{eq:scatteringwf2}
\end{align}
We first express both the plane wave and the scattered wave in spherical
wave basis. Using the first spherical Hankel function $\mathscr{\textrm{h}}_{l}^{(1)}=\mathscr{\textrm{j}}_{l}+i\mathscr{\textrm{n}}_{l}$
to construct an outgoing wave $|\psi_{\mathrm{o}}\rangle$in spherical
wave basis one obtains 
\begin{align}
|\psi_{\mathrm{o}}\rangle & =\sum_{jj_{z}}
\frac{R_{jj_{z}}}{\sqrt{2}}
\biggl[\mathscr{\textrm{h}}_{j-1/2}^{(1)}(Er)|\phi_{j,j_{z}}^{+}\rangle%
(Er)|\phi_{j,j_{z}}^{-}\rangle\biggr].
\end{align}
This way of writing the scattered wave is known as partial wave expansion
with coefficients $R_{jj_{z}}$.
The incident plane wave is regular at the origin and thus in spherical wave basis given by 
\begin{align}
|E,\mathbf{v}\rangle & =e^{iEr\left(\mathbf{v}\cdot\hat{\bm{r}}\right)}|\mathbf{v}\rangle\label{eq:psiindef2}\\
 & =4\pi\sum_{jj_{z}}i^{j-1/2}\left\langle \phi_{j,j_{z}}^{+}(\zeta,\eta)|\mathbf{v}\right\rangle \nonumber \\
 & \quad\times\biggl[\mathscr{\textrm{j}}_{j-1/2}(Er)|\phi_{j,j_{z}}^{+}\left(\theta,\varphi\right)\rangle\nonumber \\
 & \qquad-i\mathscr{\textrm{j}}_{j+1/2}(Er)|\phi_{j,j_{z}}^{-}\left(\theta,\varphi\right)\rangle\biggr].\label{eq:planwave2}
\end{align}

At $r\rightarrow\infty$, we can further express plane and spherical
waves by taking the asymptotic forms of $\mathscr{\textrm{h}}_{l}^{(1)}(x)\rightarrow e^{i(x-\left(l+1\right)\pi/2)}/x$
and $\textrm{j}_{l}(x)\rightarrow[e^{i(x-l\pi/2)}-e^{-i(x-l\pi/2)}]/2ix$.
One obtains for the incident plane wave 
\begin{align}
|E,\mathbf{v}\rangle & \rightarrow\frac{2\pi}{iEr}\sum_{jj_{z}}\left\langle \phi_{j,j_{z}}^{+}(\zeta,\eta)|\mathbf{v}\right\rangle \biggl[e^{iEr}\left(1+\bm{\sigma}\cdot\hat{\bm{r}}\right)|\phi_{j,j_{z}}^{+}\rangle\nonumber \\
 & \quad-e^{-i\left(Er-\left(j-1/2\right)\pi\right)}\left(1-\bm{\sigma}\cdot\hat{\bm{r}}\right)|\phi_{j,j_{z}}^{+}\rangle\biggr],\label{eq:psiinforrggb2}
 \shortintertext{and likewise for the outgoing wave}
 |\psi_{\mathrm{o}}\rangle & \rightarrow\frac{e^{iEr}}{Er}\sum_{jj_{z}}i^{-j-1/2}\frac{R_{jj_{z}}}{\sqrt{2}}\left(1+\bm{\sigma}\cdot\hat{\bm{r}}\right)|\phi_{j,j_{z}}^{+}\rangle.\label{eq:psifsforrggb2}
\end{align}
Inserting Eq.~(\ref{eq:psifsforrggb2}) into Eq.~(\ref{eq:scatteringwf2})
and going back from spherical wave basis to plane wave basis yields
for $f(\mathbf{v},\bm{r})$ 
\begin{align}
f(\mathbf{v},\bm{r})=\sqrt{2}\sum_{jj_{z}}i^{-j-1/2}\frac{R_{jj_{z}}\left\langle \phi_{j,j_{z}}^{+}(\theta,\varphi)\right|\left.\mathbf{r}\right\rangle ^{*}}{E}.\label{eq:defamplitude}
\end{align}
From current conservation it is clear that Eqs.~(\ref{eq:psiinforrggb2})
and (\ref{eq:psifsforrggb2}) fix the coefficients $R_{jj_{z}}$ up
to a phase shift $\delta_{j}$, namely 
\begin{align}
R_{jj_{z}} & =-4\sqrt{2}\pi i^{j-1/2}\left\langle \phi_{j,j_{z}}^{+}(\theta',\varphi')|\mathbf{v}\right\rangle e^{i\delta_{j}}.\label{eq:defR}
\end{align}
The phase shift can only be determined by solving the Schr\"odinger
equation including the impurity potential. For a rectangular potential,
Eqs.~(\ref{eq:defamplitude}) and (\ref{eq:defR}) lead to Eq.~(\ref{scamplitude})
in the main text for $f(\mathbf{v},\bm{r})$.

The scattering cross section $\sigma$ is defined through 
\begin{align}
\sigma & =\int\mathrm{d}\Omega_{\bm{r}}\left|f(\mathbf{v},\bm{r})\right|^{2}\\
 & =\sum_{jj_{z}}\frac{\left|R_{jj_{z}}\right|^{2}}{E^{2}}\\
 & =\frac{4\pi}{E^{2}}\sum_{j}(2j+1)\sin^{2}\delta_{j}\label{eq:finalsigma}
\end{align}
Here it was used that $2\left|\left\langle \phi_{j,j_{z}}^{+}(\zeta,\eta)|\mathbf{v}\right\rangle \right|^{2}=||\phi_{j,j_{z}}^{+}\rangle|^{2}-\langle\phi_{j,j_{z}}^{+}|\phi_{j,j_{z}}^{-}\rangle$,
as well as the sum rules $\sum_{j_{z}}||\phi_{j,j_{z}}^{+}\rangle|^{2}=(2j+1)/4\pi$
and $\sum_{j_{z}}\langle\phi_{j,j_{z}}^{+}|\phi_{j,j_{z}}^{-}\rangle=0$.
Eq.~(\ref{eq:finalsigma}) directly yields Eq.~(\ref{eq:specificsigma})
once the specific phase shift $\delta_{j}$ for a rectangular potential
is inserted.

\section{Energy integrals}
\label{sec:energyints}
All integrations are done via residues. To this end, we write powers of the sine as
\begin{align}
\sin^2x&=\Re\left[\frac{1-e^{2ix}}{2}\right]\\
\sin^4x&=\Re\left[\frac{3-4e^{2ix}+e^{4ix}}{8}\right].
\end{align}
For a complex function $f(x)$ it holds for example that
\begin{align}
& 2\int\dd x \sin^2x f(x)
\nonumber\\&=
\int\dd x \tfrac{1-e^{2ix}}{2} f(x)
+\left(\int\dd x \tfrac{1-e^{2ix}}{2} f^*(x)\right)^*,
\end{align}
where the integrals on the right hand side can be evaluated directly by a contour integration enclosing the upper half-plane.

Defining $n\pi f_+=E+n\pi+is$, $-n\pi f_-=E-n\pi+is$, $c_{1}=1+E/n\pi$, $c_{2}=1+2E/3n\pi$ and $\alpha=\exp(2iE-2s)$, it is
\begin{align}
& I_4^{0}(E,s)\nonumber\\
&=
\int\frac{\mathrm{d}\epsilon}{\epsilon^2}
\left(\frac{(n\pi)^2\sin^{2}\epsilon}
{(n\pi+\epsilon)^2}\right)^2
\frac{1}{E-\epsilon+i s}
\frac{1}{E-\epsilon+i0^+}\\
&=
\frac{\pi}{8s}
\left(\frac{(2-\alpha)^2-1}{(E+is)^2f_+^4}
-\frac{(2-e^{2i E})^2-1}{E^2c_1^4}\right)
\nonumber\\&\quad+
\frac{\pi}{6c_1^3f_+^3}\Biggl[
\left(\left(4+\frac{9}{n^2\pi^2}\right)c_1^2
+\frac{9c_2}{n^2\pi^2}\right)f_+^2
\nonumber\\&\quad+
\frac{9c_1c_2}{n^2\pi^2}f_++\frac{3c_1^2}{n^2\pi^2}
\Biggr]+\frac{\pi}{2E(E+is)}
\end{align}
For $E\sim s\rightarrow 0$, it is $f_+=f_-=c_1=c_2=1$ and the expression expands to
\begin{align}
\lim_{n\rightarrow\infty}I_4^{0}(E,s)
&\approx
\frac{4\pi}{3}+\frac{10}{n^2\pi}+c_{E}E+c_{s}s,
\end{align}
where $c_E$ and $c_s$ are the complex coefficients of the linear terms. For finite $E$ or $s$ these linear terms lead to a small shift in the maximum of the self consistent $|\Im\Sigma|$, which results in a slightly different condition for resonant scattering.
The same kind of calculation can be repeated for all energy integrals. It is
\begin{align}
I_2(E,s)&=
\int\mathrm{d}\epsilon
\frac{(n\pi)^2\sin^{2}\epsilon}
{(n\pi+\epsilon)^{2}}
\frac{1}{E-\epsilon+i s}
\frac{1}{E-\epsilon-i s}
\label{eq:DOSintegral}
\\&=\pi\Biggl[\Re\left[
\frac{1-\alpha}{2s f_+^2}\right]
+\frac{1}{|f_+|^2}\Biggr],
\label{eq:DOSintegral2}
\\I_2'(E,s)&=
\int\mathrm{d}\epsilon
\frac{(n\pi)^2\sin^{2}\epsilon}
{(n\pi+\epsilon)(n\pi-\epsilon)}
\nonumber\\&\quad\times
\frac{1}{E+\epsilon+i s}
\frac{1}{E-\epsilon-i s}\\
&=\frac{\pi}{2E}
\Im\left[\frac{1-\alpha}{f_+f_-}\right],
\end{align}
with the limits
\begin{align}
I_2(E,s)
&=2\pi-\pi s+\frac{\pi E^2}{s}-\frac{6 E}{n}\\
I_2'(E,s)
&=-\pi+2\pi s.
\end{align}
The integrals of the ladder element evaluate to
\begin{align}
I_4(E,s)&=
\int\frac{\mathrm{d}\epsilon}{\epsilon^2}
\left(\frac{(n\pi)^2\sin^{2}\epsilon}
{(n\pi+\epsilon)^{2}}\right)^2
\frac{1}{E-\epsilon+i s}
\frac{1}{E-\epsilon-i s}\\
&=\frac{\pi}{2(E^2+s^2)}
+\pi\Re\left[\frac{(2-\alpha)^2-1}{8s(E+is)^2f_+^4}\right]
\nonumber\\&\quad+
\frac{\pi}{6|f_+|^6}\Biggl[
\left(4+\frac{9}{n^2\pi^2}\right)|f_+|^4-\frac{12s^2}{n^4\pi^4}
\nonumber\\&\quad+
3\left(\frac{4E}{n^3\pi^3}+\frac{7}{n^2\pi^2}\right)|f_+|^2\Biggr],
\\
I_4'(E,s)&=
\int\frac{\mathrm{d}\epsilon}{\epsilon^2}
\left(\frac{(n\pi)^2\sin^{2}\epsilon}
{(n\pi+\epsilon)(n\pi-\epsilon)}\right)^2
\nonumber\\&\quad\times
\frac{1}{E+\epsilon+i s}
\frac{1}{E-\epsilon-i s}\\
&=\frac{\pi}{2(E^2+s^2)}
+\pi\Im\left[\frac{(2-\alpha)^2-1}{8E(E+is)^2f_+^2f_-^2}\right]
\nonumber\\&\quad
\frac{1}{8n^2\pi E}\Re\left[\frac{1}{f_+}-\frac{1}{f_-}\right],
\end{align}
which can be expanded to obtain
\begin{align}
I_4(E,s)
&=\frac{4\pi}{3}+\frac{10}{n^2\pi}-\pi s+\frac{\pi E^2}{s}\\
I_4'(E,s)
&=-\frac{2\pi}{3}-\frac{5}{4n^2\pi}+2\pi s.
\end{align}
We note that unlike for $I_2$, the emergence of a singular term $E^2/s$ in the expansion of $I_4$ does not signify any physical divergence. To see this, we remind the reader that $s$ is essentially the imaginary part of the self-energy taken at zero, $\Im\Sigma^R(0,0)=n_i\Im\gamma=-s$. For strong scatterers, $s\sim \sqrt{n_i}$, while in the perturbative limit $s\sim E^2$. As $I_4$ only occurs in the combination $n_i|\gamma|^2I_4$, there is no singularity.

We refrain from presenting the more complicated integrals explicitly, as they are straightforwardly evaluated in the same manner.
\begin{align}
I_4''(E,s)&=
\int\frac{\mathrm{d}\epsilon}{\epsilon^2}
\left(\frac{(n\pi)^2\sin^{2}\epsilon}
{(n\pi+\epsilon)^{2}}\right)^2
\nonumber\\&\quad\times
\left(\frac{1}{E-\epsilon+i s}
\frac{1}{E-\epsilon-i s}\right)^2\\
&\approx \frac{\pi}{2}\left(\frac{E^2}{s^3}+\frac{1}{s}\right)
\end{align}

\begin{align}
I_4'''(E,s)&=
\int\frac{\mathrm{d}\epsilon}{\epsilon^2}
\left(\frac{(n\pi)^2\sin^{2}\epsilon}
{(n\pi+\epsilon)(n\pi-\epsilon)}\right)^2
\nonumber\\&\quad\times
\frac{1}{E-\epsilon+i s}
\frac{1}{E-\epsilon-i s}
\nonumber\\&\quad\times
\frac{1}{E+\epsilon+i s}
\frac{1}{E+\epsilon-i s}\\
&\approx \frac{\pi}{2s}
\end{align}

The contributions to $\Pi^{RR}$ and $\Pi^{AA}$ are complex conjugate (e.g. $I_2^{RR}=(I_2^{AA})^*$). It is
\begin{align}
I_2^{RR}&=\frac{n^2\pi^3}{f_+^2}
\left[1+\alpha-i\frac{1-\alpha}{f_+}\right]\\
&\approx 2\pi+\left(2\pi i-\tfrac{6}{n}\right) (E+is)\\
I_2'^{RR}&=i\frac{n^2\pi^3(1-\alpha)}{2(E+is)f_+f_-}\\
&\approx -\pi-\pi i (E+is),
\end{align}
and finally
\begin{align}
I_4^{RR}&\approx \frac{4\pi}{3}+\frac{10}{n^2\pi}\\
I_4'^{RR}&\approx -\frac{2\pi}{3}-\frac{5}{4n^2\pi}.
\end{align}

\section{charge conservation}
\label{app:ward}
We present an alternative proof that the particle-hole ladder implements charge conservation even when using the self-consistent T-matrix approximation. The goal is to show that asymptotically for $E\approx0$ it holds
\begin{align}
3P&=\frac{n_i|\gamma|^2}{4\pi^2}\int \epsilon^2\dd \epsilon t^4(\epsilon) G^A(E,\epsilon)G^R(E,\epsilon)\\
&= 1,
\end{align}
while not explicitly inserting the self-consistent solution. We begin with the relation
\begin{align}
\int\epsilon^2\dd \epsilon t^4(\epsilon)& G^A(E,\epsilon)G^R(E,\epsilon)
\nonumber\\
&=\int\epsilon^2\dd \epsilon t^4(\epsilon)
\frac{G^R(E,\epsilon)-G^A(E,\epsilon)}
{\Sigma^R(E,\epsilon)-\Sigma^A(E,\epsilon)}
\end{align}
Inserting $\Sigma^R(E,\epsilon)-\Sigma^A(E,\epsilon)=n_i(\gamma^R-\gamma^A)t^2(\epsilon)$, this becomes
\begin{align}
n_i\gamma^R\gamma^A
&\int\epsilon^2\dd \epsilon t^4(\epsilon) G^A(E,\epsilon)G^R(E,\epsilon)
\nonumber\\
&=\int\epsilon^2\dd \epsilon t^2(\epsilon)
\frac{G^R(E,\epsilon)-G^A(E,\epsilon)}
{\frac{1}{\gamma^A}-\frac{1}{\gamma^R}}.
\label{eq:ward1}
\end{align}
For the bare Green's functions it holds that
\begin{align}
\int\epsilon^2\dd \epsilon t^2(\epsilon)(G^A_0(E,\epsilon)-G^R_0(E,\epsilon))&=
2\pi i\sin^2 E.
\end{align}
Adding a zero to Eq.~(\ref{eq:ward1}), and recalling that the self-consistent self energy was obtained by solving
$\int\epsilon^2\dd \epsilon t^2(G^R-G^R_0)=4\pi^2({\gamma^R_0}^{-1}-{\gamma^R}^{-1})$, one arrives at
\begin{align}
&  \frac{n_i|\gamma|^2}{4\pi^2}\int\epsilon^2\dd \epsilon t^4(\epsilon) G^A(E,\epsilon)G^R(E,\epsilon)
=\frac{1}{4\pi^2}\frac{1}{\frac{1}{\gamma^A}-\frac{1}{\gamma^R}}
\nonumber\\&
\times\biggl(-2\pi i\sin^2 E+\int\epsilon^2\dd \epsilon t^2(\epsilon)(G^R(E,\epsilon)-G^R_0(E,\epsilon))
\nonumber\\&\quad
-\int\epsilon^2\dd \epsilon t^2(\epsilon)(G^A(E,\epsilon)-G^A_0(E,\epsilon))
\biggr)\\
&=1+\frac{\frac{1}{\gamma^R_0}-\frac{1}{\gamma^A_0}-i\frac{\sin^2 E}{2\pi}}{\frac{1}{\gamma^A}-\frac{1}{\gamma^R}}.
\end{align}
The bare singular amplitude is $\gamma^R_0=-\frac{4\pi}{\delta-iE^2}$, which means that
\begin{align}
3P&=1+\frac{\mathcal{O}(\delta^2,E^4)}{\frac{1}{\gamma^A}-\frac{1}{\gamma^R}}.
\end{align}
The inverse scattering amplitude goes to zero no faster than $\sim E^2$, therefore charge is conserved at least to order $\mathcal{O}(E^2)$.


\begin{thebibliography}{55}%
\makeatletter
\providecommand \@ifxundefined [1]{%
 \@ifx{#1\undefined}
}%
\providecommand \@ifnum [1]{%
 \ifnum #1\expandafter \@firstoftwo
 \else \expandafter \@secondoftwo
 \fi
}%
\providecommand \@ifx [1]{%
 \ifx #1\expandafter \@firstoftwo
 \else \expandafter \@secondoftwo
 \fi
}%
\providecommand \natexlab [1]{#1}%
\providecommand \enquote  [1]{``#1''}%
\providecommand \bibnamefont  [1]{#1}%
\providecommand \bibfnamefont [1]{#1}%
\providecommand \citenamefont [1]{#1}%
\providecommand \href@noop [0]{\@secondoftwo}%
\providecommand \href [0]{\begingroup \@sanitize@url \@href}%
\providecommand \@href[1]{\@@startlink{#1}\@@href}%
\providecommand \@@href[1]{\endgroup#1\@@endlink}%
\providecommand \@sanitize@url [0]{\catcode `\\12\catcode `\$12\catcode
  `\&12\catcode `\#12\catcode `\^12\catcode `\_12\catcode `\%12\relax}%
\providecommand \@@startlink[1]{}%
\providecommand \@@endlink[0]{}%
\providecommand \url  [0]{\begingroup\@sanitize@url \@url }%
\providecommand \@url [1]{\endgroup\@href {#1}{\urlprefix }}%
\providecommand \urlprefix  [0]{URL }%
\providecommand \Eprint [0]{\href }%
\providecommand \doibase [0]{http://dx.doi.org/}%
\providecommand \selectlanguage [0]{\@gobble}%
\providecommand \bibinfo  [0]{\@secondoftwo}%
\providecommand \bibfield  [0]{\@secondoftwo}%
\providecommand \translation [1]{[#1]}%
\providecommand \BibitemOpen [0]{}%
\providecommand \bibitemStop [0]{}%
\providecommand \bibitemNoStop [0]{.\EOS\space}%
\providecommand \EOS [0]{\spacefactor3000\relax}%
\providecommand \BibitemShut  [1]{\csname bibitem#1\endcsname}%
\let\auto@bib@innerbib\@empty
%</preamble>
\bibitem [{\citenamefont {{Hasan}}\ and\ \citenamefont
  {{Moore}}(2011)}]{Hasan2011}%
  \BibitemOpen
  \bibfield  {author} {\bibinfo {author} {\bibfnamefont {M.~Z.}\ \bibnamefont
  {{Hasan}}}\ and\ \bibinfo {author} {\bibfnamefont {J.~E.}\ \bibnamefont
  {{Moore}}},\ }\bibfield  {title} {\enquote {\bibinfo {title}
  {{Three-Dimensional Topological Insulators}},}\ }\href {\doibase
  10.1146/annurev-conmatphys-062910-140432} {\bibfield  {journal} {\bibinfo
  {journal} {\annrevcond}\ }\textbf {\bibinfo {volume} {2}},\ \bibinfo {pages}
  {55--78} (\bibinfo {year} {2011})},\ \Eprint {http://arxiv.org/abs/1011.5462}
  {arXiv:1011.5462 [cond-mat.str-el]} \BibitemShut {NoStop}%
\bibitem [{\citenamefont {{Qi}}\ and\ \citenamefont {{Zhang}}(2011)}]{Qi2011}%
  \BibitemOpen
  \bibfield  {author} {\bibinfo {author} {\bibfnamefont {X.-L.}\ \bibnamefont
  {{Qi}}}\ and\ \bibinfo {author} {\bibfnamefont {S.-C.}\ \bibnamefont
  {{Zhang}}},\ }\bibfield  {title} {\enquote {\bibinfo {title} {{Topological
  insulators and superconductors}},}\ }\href {\doibase
  10.1103/RevModPhys.83.1057} {\bibfield  {journal} {\bibinfo  {journal}
  {\rmp}\ }\textbf {\bibinfo {volume} {83}},\ \bibinfo {pages} {1057--1110}
  (\bibinfo {year} {2011})},\ \Eprint {http://arxiv.org/abs/1008.2026}
  {arXiv:1008.2026 [cond-mat.mes-hall]} \BibitemShut {NoStop}%
\bibitem [{\citenamefont {{Matsuura}}\ \emph {et~al.}(2013)\citenamefont
  {{Matsuura}}, \citenamefont {{Chang}}, \citenamefont {{Schnyder}},\ and\
  \citenamefont {{Ryu}}}]{Matsuura2013}%
  \BibitemOpen
  \bibfield  {author} {\bibinfo {author} {\bibfnamefont {S.}~\bibnamefont
  {{Matsuura}}}, \bibinfo {author} {\bibfnamefont {P.-Y.}\ \bibnamefont
  {{Chang}}}, \bibinfo {author} {\bibfnamefont {A.~P.}\ \bibnamefont
  {{Schnyder}}}, \ and\ \bibinfo {author} {\bibfnamefont {S.}~\bibnamefont
  {{Ryu}}},\ }\bibfield  {title} {\enquote {\bibinfo {title} {{Protected
  boundary states in gapless topological phases}},}\ }\href {\doibase
  10.1088/1367-2630/15/6/065001} {\bibfield  {journal} {\bibinfo  {journal}
  {\njp}\ }\textbf {\bibinfo {volume} {15}},\ \bibinfo {eid} {065001} (\bibinfo
  {year} {2013})},\ \Eprint {http://arxiv.org/abs/1212.2673} {arXiv:1212.2673
  [cond-mat.supr-con]} \BibitemShut {NoStop}%
\bibitem [{\citenamefont {{Turner}}\ and\ \citenamefont
  {{Vishwanath}}(2013)}]{Turner2013}%
  \BibitemOpen
  \bibfield  {author} {\bibinfo {author} {\bibfnamefont {A.~M.}\ \bibnamefont
  {{Turner}}}\ and\ \bibinfo {author} {\bibfnamefont {A.}~\bibnamefont
  {{Vishwanath}}},\ }\bibfield  {title} {\enquote {\bibinfo {title} {{Beyond
  Band Insulators: Topology of Semi-metals and Interacting Phases}},}\
  }\href@noop {} {\bibfield  {journal} {\bibinfo  {journal} {\arx}\ } (\bibinfo
  {year} {2013})},\ \Eprint {http://arxiv.org/abs/1301.0330} {arXiv:1301.0330
  [cond-mat.str-el]} \BibitemShut {NoStop}%
\bibitem [{\citenamefont {{Vafek}}\ and\ \citenamefont
  {{Vishwanath}}(2014)}]{Vafek2014}%
  \BibitemOpen
  \bibfield  {author} {\bibinfo {author} {\bibfnamefont {O.}~\bibnamefont
  {{Vafek}}}\ and\ \bibinfo {author} {\bibfnamefont {A.}~\bibnamefont
  {{Vishwanath}}},\ }\bibfield  {title} {\enquote {\bibinfo {title} {{Dirac
  Fermions in Solids - from High Tc cuprates and Graphene to Topological
  Insulators and Weyl Semimetals}},}\ }\href@noop {} {\bibfield  {journal}
  {\bibinfo  {journal} {\annrevcond}\ } (\bibinfo {year} {2014})},\ \Eprint
  {http://arxiv.org/abs/1306.2272} {arXiv:1306.2272 [cond-mat.mes-hall]}
  \BibitemShut {NoStop}%
\bibitem [{\citenamefont {{Metlitski}}\ \emph {et~al.}(2014)\citenamefont
  {{Metlitski}}, \citenamefont {{Fidkowski}}, \citenamefont {{Chen}},\ and\
  \citenamefont {{Vishwanath}}}]{Metlitski2014}%
  \BibitemOpen
  \bibfield  {author} {\bibinfo {author} {\bibfnamefont {M.~A.}\ \bibnamefont
  {{Metlitski}}}, \bibinfo {author} {\bibfnamefont {L.}~\bibnamefont
  {{Fidkowski}}}, \bibinfo {author} {\bibfnamefont {X.}~\bibnamefont {{Chen}}},
  \ and\ \bibinfo {author} {\bibfnamefont {A.}~\bibnamefont {{Vishwanath}}},\
  }\bibfield  {title} {\enquote {\bibinfo {title} {{Interaction effects on 3D
  topological superconductors: surface topological order from vortex
  condensation, the 16 fold way and fermionic Kramers doublets}},}\ }\href@noop
  {} {\bibfield  {journal} {\bibinfo  {journal} {\arx}\ } (\bibinfo {year}
  {2014})},\ \Eprint {http://arxiv.org/abs/1406.3032} {arXiv:1406.3032
  [cond-mat.str-el]} \BibitemShut {NoStop}%
\bibitem [{\citenamefont {{Burkov}}\ and\ \citenamefont
  {{Balents}}(2011)}]{Burkov2011}%
  \BibitemOpen
  \bibfield  {author} {\bibinfo {author} {\bibfnamefont {A.~A.}\ \bibnamefont
  {{Burkov}}}\ and\ \bibinfo {author} {\bibfnamefont {L.}~\bibnamefont
  {{Balents}}},\ }\bibfield  {title} {\enquote {\bibinfo {title} {{Weyl
  Semimetal in a Topological Insulator Multilayer}},}\ }\href {\doibase
  10.1103/PhysRevLett.107.127205} {\bibfield  {journal} {\bibinfo  {journal}
  {\prl}\ }\textbf {\bibinfo {volume} {107}},\ \bibinfo {eid} {127205}
  (\bibinfo {year} {2011})},\ \Eprint {http://arxiv.org/abs/1105.5138}
  {arXiv:1105.5138 [cond-mat.mes-hall]} \BibitemShut {NoStop}%
\bibitem [{\citenamefont {{Wan}}\ \emph {et~al.}(2011)\citenamefont {{Wan}},
  \citenamefont {{Turner}}, \citenamefont {{Vishwanath}},\ and\ \citenamefont
  {{Savrasov}}}]{Wan2011}%
  \BibitemOpen
  \bibfield  {author} {\bibinfo {author} {\bibfnamefont {X.}~\bibnamefont
  {{Wan}}}, \bibinfo {author} {\bibfnamefont {A.~M.}\ \bibnamefont {{Turner}}},
  \bibinfo {author} {\bibfnamefont {A.}~\bibnamefont {{Vishwanath}}}, \ and\
  \bibinfo {author} {\bibfnamefont {S.~Y.}\ \bibnamefont {{Savrasov}}},\
  }\bibfield  {title} {\enquote {\bibinfo {title} {{Topological semimetal and
  Fermi-arc surface states in the electronic structure of pyrochlore
  iridates}},}\ }\href {\doibase 10.1103/PhysRevB.83.205101} {\bibfield
  {journal} {\bibinfo  {journal} {\prb}\ }\textbf {\bibinfo {volume} {83}},\
  \bibinfo {eid} {205101} (\bibinfo {year} {2011})},\ \Eprint
  {http://arxiv.org/abs/1007.0016} {arXiv:1007.0016 [cond-mat.str-el]}
  \BibitemShut {NoStop}%
\bibitem [{\citenamefont {{Nielsen}}\ and\ \citenamefont
  {{Ninomiya}}(1983)}]{Nielsen1983}%
  \BibitemOpen
  \bibfield  {author} {\bibinfo {author} {\bibfnamefont {H.~B.}\ \bibnamefont
  {{Nielsen}}}\ and\ \bibinfo {author} {\bibfnamefont {M.}~\bibnamefont
  {{Ninomiya}}},\ }\bibfield  {title} {\enquote {\bibinfo {title} {{The
  Adler-Bell-Jackiw anomaly and Weyl fermions in a crystal}},}\ }\href
  {\doibase 10.1016/0370-2693(83)91529-0} {\bibfield  {journal} {\bibinfo
  {journal} {Physics Letters B}\ }\textbf {\bibinfo {volume} {130}},\ \bibinfo
  {pages} {389--396} (\bibinfo {year} {1983})}\BibitemShut {NoStop}%
\bibitem [{\citenamefont {{Son}}\ and\ \citenamefont
  {{Yamamoto}}(2012)}]{Son2012}%
  \BibitemOpen
  \bibfield  {author} {\bibinfo {author} {\bibfnamefont {D.~T.}\ \bibnamefont
  {{Son}}}\ and\ \bibinfo {author} {\bibfnamefont {N.}~\bibnamefont
  {{Yamamoto}}},\ }\bibfield  {title} {\enquote {\bibinfo {title} {{Berry
  Curvature, Triangle Anomalies, and the Chiral Magnetic Effect in Fermi
  Liquids}},}\ }\href {\doibase 10.1103/PhysRevLett.109.181602} {\bibfield
  {journal} {\bibinfo  {journal} {\prl}\ }\textbf {\bibinfo {volume} {109}},\
  \bibinfo {eid} {181602} (\bibinfo {year} {2012})},\ \Eprint
  {http://arxiv.org/abs/1203.2697} {arXiv:1203.2697 [cond-mat.mes-hall]}
  \BibitemShut {NoStop}%
\bibitem [{\citenamefont {{Chen}}\ \emph {et~al.}(2013)\citenamefont {{Chen}},
  \citenamefont {{Wu}},\ and\ \citenamefont {{Burkov}}}]{Chen2013}%
  \BibitemOpen
  \bibfield  {author} {\bibinfo {author} {\bibfnamefont {Y.}~\bibnamefont
  {{Chen}}}, \bibinfo {author} {\bibfnamefont {S.}~\bibnamefont {{Wu}}}, \ and\
  \bibinfo {author} {\bibfnamefont {A.~A.}\ \bibnamefont {{Burkov}}},\
  }\bibfield  {title} {\enquote {\bibinfo {title} {{Axion response in Weyl
  semimetals}},}\ }\href {\doibase 10.1103/PhysRevB.88.125105} {\bibfield
  {journal} {\bibinfo  {journal} {\prb}\ }\textbf {\bibinfo {volume} {88}},\
  \bibinfo {eid} {125105} (\bibinfo {year} {2013})},\ \Eprint
  {http://arxiv.org/abs/1306.5344} {arXiv:1306.5344 [cond-mat.mes-hall]}
  \BibitemShut {NoStop}%
\bibitem [{\citenamefont {{Son}}\ and\ \citenamefont
  {{Spivak}}(2013)}]{Son2013}%
  \BibitemOpen
  \bibfield  {author} {\bibinfo {author} {\bibfnamefont {D.~T.}\ \bibnamefont
  {{Son}}}\ and\ \bibinfo {author} {\bibfnamefont {B.~Z.}\ \bibnamefont
  {{Spivak}}},\ }\bibfield  {title} {\enquote {\bibinfo {title} {{Chiral
  anomaly and classical negative magnetoresistance of Weyl metals}},}\ }\href
  {\doibase 10.1103/PhysRevB.88.104412} {\bibfield  {journal} {\bibinfo
  {journal} {\prb}\ }\textbf {\bibinfo {volume} {88}},\ \bibinfo {eid} {104412}
  (\bibinfo {year} {2013})},\ \Eprint {http://arxiv.org/abs/1206.1627}
  {arXiv:1206.1627 [cond-mat.mes-hall]} \BibitemShut {NoStop}%
\bibitem [{\citenamefont {{Burkov}}(2014)}]{Burkov2014}%
  \BibitemOpen
  \bibfield  {author} {\bibinfo {author} {\bibfnamefont {A.~A.}\ \bibnamefont
  {{Burkov}}},\ }\bibfield  {title} {\enquote {\bibinfo {title} {{Chiral
  Anomaly and Diffusive Magnetotransport in Weyl Metals}},}\ }\href {\doibase
  10.1103/PhysRevLett.113.247203} {\bibfield  {journal} {\bibinfo  {journal}
  {\prl}\ }\textbf {\bibinfo {volume} {113}},\ \bibinfo {eid} {247203}
  (\bibinfo {year} {2014})},\ \Eprint {http://arxiv.org/abs/1409.0013}
  {arXiv:1409.0013 [cond-mat.mes-hall]} \BibitemShut {NoStop}%
\bibitem [{\citenamefont {{Gorbar}}\ \emph {et~al.}(2014)\citenamefont
  {{Gorbar}}, \citenamefont {{Miransky}},\ and\ \citenamefont
  {{Shovkovy}}}]{Gorbar2014}%
  \BibitemOpen
  \bibfield  {author} {\bibinfo {author} {\bibfnamefont {E.~V.}\ \bibnamefont
  {{Gorbar}}}, \bibinfo {author} {\bibfnamefont {V.~A.}\ \bibnamefont
  {{Miransky}}}, \ and\ \bibinfo {author} {\bibfnamefont {I.~A.}\ \bibnamefont
  {{Shovkovy}}},\ }\bibfield  {title} {\enquote {\bibinfo {title} {{Chiral
  anomaly, dimensional reduction, and magnetoresistivity of Weyl and Dirac
  semimetals}},}\ }\href {\doibase 10.1103/PhysRevB.89.085126} {\bibfield
  {journal} {\bibinfo  {journal} {\prb}\ }\textbf {\bibinfo {volume} {89}},\
  \bibinfo {eid} {085126} (\bibinfo {year} {2014})},\ \Eprint
  {http://arxiv.org/abs/1312.0027} {arXiv:1312.0027 [cond-mat.mes-hall]}
  \BibitemShut {NoStop}%
\bibitem [{\citenamefont {{Baum}}\ \emph {et~al.}(2015)\citenamefont {{Baum}},
  \citenamefont {{Berg}}, \citenamefont {{Parameswaran}},\ and\ \citenamefont
  {{Stern}}}]{Baum2015}%
  \BibitemOpen
  \bibfield  {author} {\bibinfo {author} {\bibfnamefont {Y.}~\bibnamefont
  {{Baum}}}, \bibinfo {author} {\bibfnamefont {E.}~\bibnamefont {{Berg}}},
  \bibinfo {author} {\bibfnamefont {S.~A.}\ \bibnamefont {{Parameswaran}}}, \
  and\ \bibinfo {author} {\bibfnamefont {A.}~\bibnamefont {{Stern}}},\
  }\bibfield  {title} {\enquote {\bibinfo {title} {{Current at a Distance and
  Resonant Transparency in Weyl Semimetals}},}\ }\href {\doibase
  10.1103/PhysRevX.5.041046} {\bibfield  {journal} {\bibinfo  {journal} {\prx}\
  }\textbf {\bibinfo {volume} {5}},\ \bibinfo {eid} {041046} (\bibinfo {year}
  {2015})},\ \Eprint {http://arxiv.org/abs/1508.03047} {arXiv:1508.03047
  [cond-mat.mes-hall]} \BibitemShut {NoStop}%
\bibitem [{\citenamefont {{Xu}}\ \emph
  {et~al.}(2015{\natexlab{a}})\citenamefont {{Xu}}, \citenamefont
  {{Belopolski}}, \citenamefont {{Alidoust}}, \citenamefont {{Neupane}},
  \citenamefont {{Bian}}, \citenamefont {{Zhang}}, \citenamefont {{Sankar}},
  \citenamefont {{Chang}}, \citenamefont {{Yuan}}, \citenamefont {{Lee}},
  \citenamefont {{Huang}}, \citenamefont {{Zheng}}, \citenamefont {{Ma}},
  \citenamefont {{Sanchez}}, \citenamefont {{Wang}}, \citenamefont {{Bansil}},
  \citenamefont {{Chou}}, \citenamefont {{Shibayev}}, \citenamefont {{Lin}},
  \citenamefont {{Jia}},\ and\ \citenamefont {{Hasan}}}]{Xu2015}%
  \BibitemOpen
  \bibfield  {author} {\bibinfo {author} {\bibfnamefont {S.-Y.}\ \bibnamefont
  {{Xu}}}, \bibinfo {author} {\bibfnamefont {I.}~\bibnamefont {{Belopolski}}},
  \bibinfo {author} {\bibfnamefont {N.}~\bibnamefont {{Alidoust}}}, \bibinfo
  {author} {\bibfnamefont {M.}~\bibnamefont {{Neupane}}}, \bibinfo {author}
  {\bibfnamefont {G.}~\bibnamefont {{Bian}}}, \bibinfo {author} {\bibfnamefont
  {C.}~\bibnamefont {{Zhang}}}, \bibinfo {author} {\bibfnamefont
  {R.}~\bibnamefont {{Sankar}}}, \bibinfo {author} {\bibfnamefont
  {G.}~\bibnamefont {{Chang}}}, \bibinfo {author} {\bibfnamefont
  {Z.}~\bibnamefont {{Yuan}}}, \bibinfo {author} {\bibfnamefont {C.-C.}\
  \bibnamefont {{Lee}}}, \bibinfo {author} {\bibfnamefont {S.-M.}\ \bibnamefont
  {{Huang}}}, \bibinfo {author} {\bibfnamefont {H.}~\bibnamefont {{Zheng}}},
  \bibinfo {author} {\bibfnamefont {J.}~\bibnamefont {{Ma}}}, \bibinfo {author}
  {\bibfnamefont {D.~S.}\ \bibnamefont {{Sanchez}}}, \bibinfo {author}
  {\bibfnamefont {B.}~\bibnamefont {{Wang}}}, \bibinfo {author} {\bibfnamefont
  {A.}~\bibnamefont {{Bansil}}}, \bibinfo {author} {\bibfnamefont
  {F.}~\bibnamefont {{Chou}}}, \bibinfo {author} {\bibfnamefont {P.~P.}\
  \bibnamefont {{Shibayev}}}, \bibinfo {author} {\bibfnamefont
  {H.}~\bibnamefont {{Lin}}}, \bibinfo {author} {\bibfnamefont
  {S.}~\bibnamefont {{Jia}}}, \ and\ \bibinfo {author} {\bibfnamefont {M.~Z.}\
  \bibnamefont {{Hasan}}},\ }\bibfield  {title} {\enquote {\bibinfo {title}
  {{Discovery of a Weyl fermion semimetal and topological Fermi arcs}},}\
  }\href {\doibase 10.1126/science.aaa9297} {\bibfield  {journal} {\bibinfo
  {journal} {Science}\ }\textbf {\bibinfo {volume} {349}},\ \bibinfo {pages}
  {613--617} (\bibinfo {year} {2015}{\natexlab{a}})},\ \Eprint
  {http://arxiv.org/abs/1502.03807} {arXiv:1502.03807 [cond-mat.mes-hall]}
  \BibitemShut {NoStop}%
\bibitem [{\citenamefont {{Lv}}\ \emph
  {et~al.}(2015{\natexlab{a}})\citenamefont {{Lv}}, \citenamefont {{Weng}},
  \citenamefont {{Fu}}, \citenamefont {{Wang}}, \citenamefont {{Miao}},
  \citenamefont {{Ma}}, \citenamefont {{Richard}}, \citenamefont {{Huang}},
  \citenamefont {{Zhao}}, \citenamefont {{Chen}}, \citenamefont {{Fang}},
  \citenamefont {{Dai}}, \citenamefont {{Qian}},\ and\ \citenamefont
  {{Ding}}}]{Lv2015}%
  \BibitemOpen
  \bibfield  {author} {\bibinfo {author} {\bibfnamefont {B.~Q.}\ \bibnamefont
  {{Lv}}}, \bibinfo {author} {\bibfnamefont {H.~M.}\ \bibnamefont {{Weng}}},
  \bibinfo {author} {\bibfnamefont {B.~B.}\ \bibnamefont {{Fu}}}, \bibinfo
  {author} {\bibfnamefont {X.~P.}\ \bibnamefont {{Wang}}}, \bibinfo {author}
  {\bibfnamefont {H.}~\bibnamefont {{Miao}}}, \bibinfo {author} {\bibfnamefont
  {J.}~\bibnamefont {{Ma}}}, \bibinfo {author} {\bibfnamefont {P.}~\bibnamefont
  {{Richard}}}, \bibinfo {author} {\bibfnamefont {X.~C.}\ \bibnamefont
  {{Huang}}}, \bibinfo {author} {\bibfnamefont {L.~X.}\ \bibnamefont {{Zhao}}},
  \bibinfo {author} {\bibfnamefont {G.~F.}\ \bibnamefont {{Chen}}}, \bibinfo
  {author} {\bibfnamefont {Z.}~\bibnamefont {{Fang}}}, \bibinfo {author}
  {\bibfnamefont {X.}~\bibnamefont {{Dai}}}, \bibinfo {author} {\bibfnamefont
  {T.}~\bibnamefont {{Qian}}}, \ and\ \bibinfo {author} {\bibfnamefont
  {H.}~\bibnamefont {{Ding}}},\ }\bibfield  {title} {\enquote {\bibinfo {title}
  {{Experimental Discovery of Weyl Semimetal TaAs}},}\ }\href {\doibase
  10.1103/PhysRevX.5.031013} {\bibfield  {journal} {\bibinfo  {journal} {\prx}\
  }\textbf {\bibinfo {volume} {5}},\ \bibinfo {eid} {031013} (\bibinfo {year}
  {2015}{\natexlab{a}})},\ \Eprint {http://arxiv.org/abs/1502.04684}
  {arXiv:1502.04684 [cond-mat.mtrl-sci]} \BibitemShut {NoStop}%
\bibitem [{\citenamefont {{Xu}}\ \emph
  {et~al.}(2015{\natexlab{b}})\citenamefont {{Xu}}, \citenamefont {{Alidoust}},
  \citenamefont {{Belopolski}}, \citenamefont {{Yuan}}, \citenamefont {{Bian}},
  \citenamefont {{Chang}}, \citenamefont {{Zheng}}, \citenamefont {{Strocov}},
  \citenamefont {{Sanchez}}, \citenamefont {{Chang}}, \citenamefont {{Zhang}},
  \citenamefont {{Mou}}, \citenamefont {{Wu}}, \citenamefont {{Huang}},
  \citenamefont {{Lee}}, \citenamefont {{Huang}}, \citenamefont {{Wang}},
  \citenamefont {{Bansil}}, \citenamefont {{Jeng}}, \citenamefont {{Neupert}},
  \citenamefont {{Kaminski}}, \citenamefont {{Lin}}, \citenamefont {{Jia}},\
  and\ \citenamefont {{Zahid Hasan}}}]{Xu2015a}%
  \BibitemOpen
  \bibfield  {author} {\bibinfo {author} {\bibfnamefont {S.-Y.}\ \bibnamefont
  {{Xu}}}, \bibinfo {author} {\bibfnamefont {N.}~\bibnamefont {{Alidoust}}},
  \bibinfo {author} {\bibfnamefont {I.}~\bibnamefont {{Belopolski}}}, \bibinfo
  {author} {\bibfnamefont {Z.}~\bibnamefont {{Yuan}}}, \bibinfo {author}
  {\bibfnamefont {G.}~\bibnamefont {{Bian}}}, \bibinfo {author} {\bibfnamefont
  {T.-R.}\ \bibnamefont {{Chang}}}, \bibinfo {author} {\bibfnamefont
  {H.}~\bibnamefont {{Zheng}}}, \bibinfo {author} {\bibfnamefont {V.~N.}\
  \bibnamefont {{Strocov}}}, \bibinfo {author} {\bibfnamefont {D.~S.}\
  \bibnamefont {{Sanchez}}}, \bibinfo {author} {\bibfnamefont {G.}~\bibnamefont
  {{Chang}}}, \bibinfo {author} {\bibfnamefont {C.}~\bibnamefont {{Zhang}}},
  \bibinfo {author} {\bibfnamefont {D.}~\bibnamefont {{Mou}}}, \bibinfo
  {author} {\bibfnamefont {Y.}~\bibnamefont {{Wu}}}, \bibinfo {author}
  {\bibfnamefont {L.}~\bibnamefont {{Huang}}}, \bibinfo {author} {\bibfnamefont
  {C.-C.}\ \bibnamefont {{Lee}}}, \bibinfo {author} {\bibfnamefont {S.-M.}\
  \bibnamefont {{Huang}}}, \bibinfo {author} {\bibfnamefont {B.}~\bibnamefont
  {{Wang}}}, \bibinfo {author} {\bibfnamefont {A.}~\bibnamefont {{Bansil}}},
  \bibinfo {author} {\bibfnamefont {H.-T.}\ \bibnamefont {{Jeng}}}, \bibinfo
  {author} {\bibfnamefont {T.}~\bibnamefont {{Neupert}}}, \bibinfo {author}
  {\bibfnamefont {A.}~\bibnamefont {{Kaminski}}}, \bibinfo {author}
  {\bibfnamefont {H.}~\bibnamefont {{Lin}}}, \bibinfo {author} {\bibfnamefont
  {S.}~\bibnamefont {{Jia}}}, \ and\ \bibinfo {author} {\bibfnamefont
  {M.}~\bibnamefont {{Zahid Hasan}}},\ }\bibfield  {title} {\enquote {\bibinfo
  {title} {{Discovery of a Weyl fermion state with Fermi arcs in niobium
  arsenide}},}\ }\href {\doibase 10.1038/nphys3437} {\bibfield  {journal}
  {\bibinfo  {journal} {Nat. Phys.}\ }\textbf {\bibinfo {volume} {11}},\
  \bibinfo {pages} {748--754} (\bibinfo {year}
  {2015}{\natexlab{b}})}\BibitemShut {NoStop}%
\bibitem [{\citenamefont {{Lv}}\ \emph
  {et~al.}(2015{\natexlab{b}})\citenamefont {{Lv}}, \citenamefont {{Muff}},
  \citenamefont {{Qian}}, \citenamefont {{Song}}, \citenamefont {{Nie}},
  \citenamefont {{Xu}}, \citenamefont {{Richard}}, \citenamefont {{Matt}},
  \citenamefont {{Plumb}}, \citenamefont {{Zhao}}, \citenamefont {{Chen}},
  \citenamefont {{Fang}}, \citenamefont {{Dai}}, \citenamefont {{Dil}},
  \citenamefont {{Mesot}}, \citenamefont {{Shi}}, \citenamefont {{Weng}},\ and\
  \citenamefont {{Ding}}}]{Lv2015a}%
  \BibitemOpen
  \bibfield  {author} {\bibinfo {author} {\bibfnamefont {B.~Q.}\ \bibnamefont
  {{Lv}}}, \bibinfo {author} {\bibfnamefont {S.}~\bibnamefont {{Muff}}},
  \bibinfo {author} {\bibfnamefont {T.}~\bibnamefont {{Qian}}}, \bibinfo
  {author} {\bibfnamefont {Z.~D.}\ \bibnamefont {{Song}}}, \bibinfo {author}
  {\bibfnamefont {S.~M.}\ \bibnamefont {{Nie}}}, \bibinfo {author}
  {\bibfnamefont {N.}~\bibnamefont {{Xu}}}, \bibinfo {author} {\bibfnamefont
  {P.}~\bibnamefont {{Richard}}}, \bibinfo {author} {\bibfnamefont {C.~E.}\
  \bibnamefont {{Matt}}}, \bibinfo {author} {\bibfnamefont {N.~C.}\
  \bibnamefont {{Plumb}}}, \bibinfo {author} {\bibfnamefont {L.~X.}\
  \bibnamefont {{Zhao}}}, \bibinfo {author} {\bibfnamefont {G.~F.}\
  \bibnamefont {{Chen}}}, \bibinfo {author} {\bibfnamefont {Z.}~\bibnamefont
  {{Fang}}}, \bibinfo {author} {\bibfnamefont {X.}~\bibnamefont {{Dai}}},
  \bibinfo {author} {\bibfnamefont {J.~H.}\ \bibnamefont {{Dil}}}, \bibinfo
  {author} {\bibfnamefont {J.}~\bibnamefont {{Mesot}}}, \bibinfo {author}
  {\bibfnamefont {M.}~\bibnamefont {{Shi}}}, \bibinfo {author} {\bibfnamefont
  {H.~M.}\ \bibnamefont {{Weng}}}, \ and\ \bibinfo {author} {\bibfnamefont
  {H.}~\bibnamefont {{Ding}}},\ }\bibfield  {title} {\enquote {\bibinfo {title}
  {{Observation of Fermi-Arc Spin Texture in TaAs}},}\ }\href {\doibase
  10.1103/PhysRevLett.115.217601} {\bibfield  {journal} {\bibinfo  {journal}
  {\prl}\ }\textbf {\bibinfo {volume} {115}},\ \bibinfo {eid} {217601}
  (\bibinfo {year} {2015}{\natexlab{b}})},\ \Eprint
  {http://arxiv.org/abs/1510.07256} {arXiv:1510.07256 [cond-mat.mtrl-sci]}
  \BibitemShut {NoStop}%
\bibitem [{\citenamefont {{Zhang}}\ \emph {et~al.}(2016)\citenamefont
  {{Zhang}}, \citenamefont {{Xu}}, \citenamefont {{Belopolski}}, \citenamefont
  {{Yuan}}, \citenamefont {{Lin}}, \citenamefont {{Tong}}, \citenamefont
  {{Bian}}, \citenamefont {{Alidoust}}, \citenamefont {{Lee}}, \citenamefont
  {{Huang}}, \citenamefont {{Chang}}, \citenamefont {{Chang}}, \citenamefont
  {{Hsu}}, \citenamefont {{Jeng}}, \citenamefont {{Neupane}}, \citenamefont
  {{Sanchez}}, \citenamefont {{Zheng}}, \citenamefont {{Wang}}, \citenamefont
  {{Lin}}, \citenamefont {{Zhang}}, \citenamefont {{Lu}}, \citenamefont
  {{Shen}}, \citenamefont {{Neupert}}, \citenamefont {{Zahid Hasan}},\ and\
  \citenamefont {{Jia}}}]{Zhang2016}%
  \BibitemOpen
  \bibfield  {author} {\bibinfo {author} {\bibfnamefont {C.-L.}\ \bibnamefont
  {{Zhang}}}, \bibinfo {author} {\bibfnamefont {S.-Y.}\ \bibnamefont {{Xu}}},
  \bibinfo {author} {\bibfnamefont {I.}~\bibnamefont {{Belopolski}}}, \bibinfo
  {author} {\bibfnamefont {Z.}~\bibnamefont {{Yuan}}}, \bibinfo {author}
  {\bibfnamefont {Z.}~\bibnamefont {{Lin}}}, \bibinfo {author} {\bibfnamefont
  {B.}~\bibnamefont {{Tong}}}, \bibinfo {author} {\bibfnamefont
  {G.}~\bibnamefont {{Bian}}}, \bibinfo {author} {\bibfnamefont
  {N.}~\bibnamefont {{Alidoust}}}, \bibinfo {author} {\bibfnamefont {C.-C.}\
  \bibnamefont {{Lee}}}, \bibinfo {author} {\bibfnamefont {S.-M.}\ \bibnamefont
  {{Huang}}}, \bibinfo {author} {\bibfnamefont {T.-R.}\ \bibnamefont
  {{Chang}}}, \bibinfo {author} {\bibfnamefont {G.}~\bibnamefont {{Chang}}},
  \bibinfo {author} {\bibfnamefont {C.-H.}\ \bibnamefont {{Hsu}}}, \bibinfo
  {author} {\bibfnamefont {H.-T.}\ \bibnamefont {{Jeng}}}, \bibinfo {author}
  {\bibfnamefont {M.}~\bibnamefont {{Neupane}}}, \bibinfo {author}
  {\bibfnamefont {D.~S.}\ \bibnamefont {{Sanchez}}}, \bibinfo {author}
  {\bibfnamefont {H.}~\bibnamefont {{Zheng}}}, \bibinfo {author} {\bibfnamefont
  {J.}~\bibnamefont {{Wang}}}, \bibinfo {author} {\bibfnamefont
  {H.}~\bibnamefont {{Lin}}}, \bibinfo {author} {\bibfnamefont
  {C.}~\bibnamefont {{Zhang}}}, \bibinfo {author} {\bibfnamefont {H.-Z.}\
  \bibnamefont {{Lu}}}, \bibinfo {author} {\bibfnamefont {S.-Q.}\ \bibnamefont
  {{Shen}}}, \bibinfo {author} {\bibfnamefont {T.}~\bibnamefont {{Neupert}}},
  \bibinfo {author} {\bibfnamefont {M.}~\bibnamefont {{Zahid Hasan}}}, \ and\
  \bibinfo {author} {\bibfnamefont {S.}~\bibnamefont {{Jia}}},\ }\bibfield
  {title} {\enquote {\bibinfo {title} {{Signatures of the Adler-Bell-Jackiw
  chiral anomaly in a Weyl fermion semimetal}},}\ }\href {\doibase
  10.1038/ncomms10735} {\bibfield  {journal} {\bibinfo  {journal} {Nat. Comm.}\
  }\textbf {\bibinfo {volume} {7}},\ \bibinfo {eid} {10735} (\bibinfo {year}
  {2016})},\ \Eprint {http://arxiv.org/abs/1601.04208} {arXiv:1601.04208
  [cond-mat.mtrl-sci]} \BibitemShut {NoStop}%
\bibitem [{\citenamefont {{Fradkin}}(1986{\natexlab{a}})}]{Fradkin1986}%
  \BibitemOpen
  \bibfield  {author} {\bibinfo {author} {\bibfnamefont {E.}~\bibnamefont
  {{Fradkin}}},\ }\bibfield  {title} {\enquote {\bibinfo {title} {{Critical
  behavior of disordered degenerate semiconductors. I. Models, symmetries, and
  formalism}},}\ }\href {\doibase 10.1103/PhysRevB.33.3257} {\bibfield
  {journal} {\bibinfo  {journal} {\prb}\ }\textbf {\bibinfo {volume} {33}},\
  \bibinfo {pages} {3257--3262} (\bibinfo {year}
  {1986}{\natexlab{a}})}\BibitemShut {NoStop}%
\bibitem [{\citenamefont {{Fradkin}}(1986{\natexlab{b}})}]{Fradkin1986a}%
  \BibitemOpen
  \bibfield  {author} {\bibinfo {author} {\bibfnamefont {E.}~\bibnamefont
  {{Fradkin}}},\ }\bibfield  {title} {\enquote {\bibinfo {title} {{Critical
  behavior of disordered degenerate semiconductors. II. Spectrum and transport
  properties in mean-field theory}},}\ }\href {\doibase
  10.1103/PhysRevB.33.3263} {\bibfield  {journal} {\bibinfo  {journal} {\prb}\
  }\textbf {\bibinfo {volume} {33}},\ \bibinfo {pages} {3263--3268} (\bibinfo
  {year} {1986}{\natexlab{b}})}\BibitemShut {NoStop}%
\bibitem [{\citenamefont {{Goswami}}\ and\ \citenamefont
  {{Chakravarty}}(2011)}]{Goswami2011}%
  \BibitemOpen
  \bibfield  {author} {\bibinfo {author} {\bibfnamefont {P.}~\bibnamefont
  {{Goswami}}}\ and\ \bibinfo {author} {\bibfnamefont {S.}~\bibnamefont
  {{Chakravarty}}},\ }\bibfield  {title} {\enquote {\bibinfo {title} {{Quantum
  Criticality between Topological and Band Insulators in 3+1 Dimensions}},}\
  }\href {\doibase 10.1103/PhysRevLett.107.196803} {\bibfield  {journal}
  {\bibinfo  {journal} {\prl}\ }\textbf {\bibinfo {volume} {107}},\ \bibinfo
  {eid} {196803} (\bibinfo {year} {2011})},\ \Eprint
  {http://arxiv.org/abs/1101.2210} {arXiv:1101.2210 [cond-mat.dis-nn]}
  \BibitemShut {NoStop}%
\bibitem [{\citenamefont {{Roy}}\ and\ \citenamefont {{Das
  Sarma}}(2014)}]{Roy2014}%
  \BibitemOpen
  \bibfield  {author} {\bibinfo {author} {\bibfnamefont {B.}~\bibnamefont
  {{Roy}}}\ and\ \bibinfo {author} {\bibfnamefont {S.}~\bibnamefont {{Das
  Sarma}}},\ }\bibfield  {title} {\enquote {\bibinfo {title} {{Diffusive
  quantum criticality in three-dimensional disordered Dirac semimetals}},}\
  }\href {\doibase 10.1103/PhysRevB.90.241112} {\bibfield  {journal} {\bibinfo
  {journal} {\prb}\ }\textbf {\bibinfo {volume} {90}},\ \bibinfo {eid} {241112}
  (\bibinfo {year} {2014})},\ \Eprint {http://arxiv.org/abs/1407.7026}
  {arXiv:1407.7026 [cond-mat.dis-nn]} \BibitemShut {NoStop}%
\bibitem [{\citenamefont {{Syzranov}}\ \emph {et~al.}(2015)\citenamefont
  {{Syzranov}}, \citenamefont {{Radzihovsky}},\ and\ \citenamefont
  {{Gurarie}}}]{Syzranov2015}%
  \BibitemOpen
  \bibfield  {author} {\bibinfo {author} {\bibfnamefont {S.~V.}\ \bibnamefont
  {{Syzranov}}}, \bibinfo {author} {\bibfnamefont {L.}~\bibnamefont
  {{Radzihovsky}}}, \ and\ \bibinfo {author} {\bibfnamefont {V.}~\bibnamefont
  {{Gurarie}}},\ }\bibfield  {title} {\enquote {\bibinfo {title} {{Critical
  Transport in Weakly Disordered Semiconductors and Semimetals}},}\ }\href
  {\doibase 10.1103/PhysRevLett.114.166601} {\bibfield  {journal} {\bibinfo
  {journal} {\prl}\ }\textbf {\bibinfo {volume} {114}},\ \bibinfo {eid}
  {166601} (\bibinfo {year} {2015})},\ \Eprint {http://arxiv.org/abs/1402.3737}
  {arXiv:1402.3737 [cond-mat.mes-hall]} \BibitemShut {NoStop}%
\bibitem [{\citenamefont {{Sbierski}}\ \emph {et~al.}(2015)\citenamefont
  {{Sbierski}}, \citenamefont {{Bergholtz}},\ and\ \citenamefont
  {{Brouwer}}}]{Sbierski2015}%
  \BibitemOpen
  \bibfield  {author} {\bibinfo {author} {\bibfnamefont {B.}~\bibnamefont
  {{Sbierski}}}, \bibinfo {author} {\bibfnamefont {E.~J.}\ \bibnamefont
  {{Bergholtz}}}, \ and\ \bibinfo {author} {\bibfnamefont {P.~W.}\ \bibnamefont
  {{Brouwer}}},\ }\bibfield  {title} {\enquote {\bibinfo {title} {{Quantum
  critical exponents for a disordered three-dimensional Weyl node}},}\ }\href
  {\doibase 10.1103/PhysRevB.92.115145} {\bibfield  {journal} {\bibinfo
  {journal} {\prb}\ }\textbf {\bibinfo {volume} {92}},\ \bibinfo {eid} {115145}
  (\bibinfo {year} {2015})},\ \Eprint {http://arxiv.org/abs/1505.07374}
  {arXiv:1505.07374 [cond-mat.mes-hall]} \BibitemShut {NoStop}%
\bibitem [{\citenamefont {{Liu}}\ \emph {et~al.}(2016)\citenamefont {{Liu}},
  \citenamefont {{Ohtsuki}},\ and\ \citenamefont {{Shindou}}}]{Liu2016}%
  \BibitemOpen
  \bibfield  {author} {\bibinfo {author} {\bibfnamefont {S.}~\bibnamefont
  {{Liu}}}, \bibinfo {author} {\bibfnamefont {T.}~\bibnamefont {{Ohtsuki}}}, \
  and\ \bibinfo {author} {\bibfnamefont {R.}~\bibnamefont {{Shindou}}},\
  }\bibfield  {title} {\enquote {\bibinfo {title} {{Effect of Disorder in a
  Three-Dimensional Layered Chern Insulator}},}\ }\href {\doibase
  10.1103/PhysRevLett.116.066401} {\bibfield  {journal} {\bibinfo  {journal}
  {\prl}\ }\textbf {\bibinfo {volume} {116}},\ \bibinfo {eid} {066401}
  (\bibinfo {year} {2016})},\ \Eprint {http://arxiv.org/abs/1507.02381}
  {arXiv:1507.02381 [cond-mat.dis-nn]} \BibitemShut {NoStop}%
\bibitem [{\citenamefont {{Syzranov}}\ \emph {et~al.}(2016)\citenamefont
  {{Syzranov}}, \citenamefont {{Ostrovsky}}, \citenamefont {{Gurarie}},\ and\
  \citenamefont {{Radzihovsky}}}]{Syzranov2016}%
  \BibitemOpen
  \bibfield  {author} {\bibinfo {author} {\bibfnamefont {S.~V.}\ \bibnamefont
  {{Syzranov}}}, \bibinfo {author} {\bibfnamefont {P.~M.}\ \bibnamefont
  {{Ostrovsky}}}, \bibinfo {author} {\bibfnamefont {V.}~\bibnamefont
  {{Gurarie}}}, \ and\ \bibinfo {author} {\bibfnamefont {L.}~\bibnamefont
  {{Radzihovsky}}},\ }\bibfield  {title} {\enquote {\bibinfo {title} {{Critical
  exponents at the unconventional disorder-driven transition in a Weyl
  semimetal}},}\ }\href {\doibase 10.1103/PhysRevB.93.155113} {\bibfield
  {journal} {\bibinfo  {journal} {\prb}\ }\textbf {\bibinfo {volume} {93}},\
  \bibinfo {eid} {155113} (\bibinfo {year} {2016})},\ \Eprint
  {http://arxiv.org/abs/1512.04130} {arXiv:1512.04130 [cond-mat.mes-hall]}
  \BibitemShut {NoStop}%
\bibitem [{\citenamefont {{Sbierski}}\ \emph {et~al.}(2014)\citenamefont
  {{Sbierski}}, \citenamefont {{Pohl}}, \citenamefont {{Bergholtz}},\ and\
  \citenamefont {{Brouwer}}}]{Sbierski2014}%
  \BibitemOpen
  \bibfield  {author} {\bibinfo {author} {\bibfnamefont {B.}~\bibnamefont
  {{Sbierski}}}, \bibinfo {author} {\bibfnamefont {G.}~\bibnamefont {{Pohl}}},
  \bibinfo {author} {\bibfnamefont {E.~J.}\ \bibnamefont {{Bergholtz}}}, \ and\
  \bibinfo {author} {\bibfnamefont {P.~W.}\ \bibnamefont {{Brouwer}}},\
  }\bibfield  {title} {\enquote {\bibinfo {title} {{Quantum Transport of
  Disordered Weyl Semimetals at the Nodal Point}},}\ }\href {\doibase
  10.1103/PhysRevLett.113.026602} {\bibfield  {journal} {\bibinfo  {journal}
  {\prl}\ }\textbf {\bibinfo {volume} {113}},\ \bibinfo {eid} {026602}
  (\bibinfo {year} {2014})},\ \Eprint {http://arxiv.org/abs/1402.6653}
  {arXiv:1402.6653 [cond-mat.mes-hall]} \BibitemShut {NoStop}%
\bibitem [{\citenamefont {{Ominato}}\ and\ \citenamefont
  {{Koshino}}(2014)}]{Ominato2014}%
  \BibitemOpen
  \bibfield  {author} {\bibinfo {author} {\bibfnamefont {Y.}~\bibnamefont
  {{Ominato}}}\ and\ \bibinfo {author} {\bibfnamefont {M.}~\bibnamefont
  {{Koshino}}},\ }\bibfield  {title} {\enquote {\bibinfo {title} {{Quantum
  transport in a three-dimensional Weyl electron system}},}\ }\href {\doibase
  10.1103/PhysRevB.89.054202} {\bibfield  {journal} {\bibinfo  {journal}
  {\prb}\ }\textbf {\bibinfo {volume} {89}},\ \bibinfo {eid} {054202} (\bibinfo
  {year} {2014})},\ \Eprint {http://arxiv.org/abs/1309.4206} {arXiv:1309.4206
  [cond-mat.dis-nn]} \BibitemShut {NoStop}%
\bibitem [{\citenamefont {{Kobayashi}}\ \emph {et~al.}(2014)\citenamefont
  {{Kobayashi}}, \citenamefont {{Ohtsuki}}, \citenamefont {{Imura}},\ and\
  \citenamefont {{Herbut}}}]{Kobayashi2014}%
  \BibitemOpen
  \bibfield  {author} {\bibinfo {author} {\bibfnamefont {K.}~\bibnamefont
  {{Kobayashi}}}, \bibinfo {author} {\bibfnamefont {T.}~\bibnamefont
  {{Ohtsuki}}}, \bibinfo {author} {\bibfnamefont {K.-I.}\ \bibnamefont
  {{Imura}}}, \ and\ \bibinfo {author} {\bibfnamefont {I.~F.}\ \bibnamefont
  {{Herbut}}},\ }\bibfield  {title} {\enquote {\bibinfo {title} {{Density of
  States Scaling at the Semimetal to Metal Transition in Three Dimensional
  Topological Insulators}},}\ }\href {\doibase 10.1103/PhysRevLett.112.016402}
  {\bibfield  {journal} {\bibinfo  {journal} {\prl}\ }\textbf {\bibinfo
  {volume} {112}},\ \bibinfo {eid} {016402} (\bibinfo {year} {2014})},\ \Eprint
  {http://arxiv.org/abs/1308.3953} {arXiv:1308.3953 [cond-mat.mes-hall]}
  \BibitemShut {NoStop}%
\bibitem [{\citenamefont {{Pixley}}\ \emph {et~al.}(2015)\citenamefont
  {{Pixley}}, \citenamefont {{Goswami}},\ and\ \citenamefont {{Das
  Sarma}}}]{Pixley2015}%
  \BibitemOpen
  \bibfield  {author} {\bibinfo {author} {\bibfnamefont {J.~H.}\ \bibnamefont
  {{Pixley}}}, \bibinfo {author} {\bibfnamefont {P.}~\bibnamefont {{Goswami}}},
  \ and\ \bibinfo {author} {\bibfnamefont {S.}~\bibnamefont {{Das Sarma}}},\
  }\bibfield  {title} {\enquote {\bibinfo {title} {{Anderson Localization and
  the Quantum Phase Diagram of Three Dimensional Disordered Dirac
  Semimetals}},}\ }\href {\doibase 10.1103/PhysRevLett.115.076601} {\bibfield
  {journal} {\bibinfo  {journal} {\prl}\ }\textbf {\bibinfo {volume} {115}},\
  \bibinfo {eid} {076601} (\bibinfo {year} {2015})},\ \Eprint
  {http://arxiv.org/abs/1502.07778} {arXiv:1502.07778 [cond-mat.dis-nn]}
  \BibitemShut {NoStop}%
\bibitem [{\citenamefont {{Pixley}}\ \emph
  {et~al.}(2016{\natexlab{a}})\citenamefont {{Pixley}}, \citenamefont
  {{Goswami}},\ and\ \citenamefont {{Das Sarma}}}]{Pixley2016}%
  \BibitemOpen
  \bibfield  {author} {\bibinfo {author} {\bibfnamefont {J.~H.}\ \bibnamefont
  {{Pixley}}}, \bibinfo {author} {\bibfnamefont {P.}~\bibnamefont {{Goswami}}},
  \ and\ \bibinfo {author} {\bibfnamefont {S.}~\bibnamefont {{Das Sarma}}},\
  }\bibfield  {title} {\enquote {\bibinfo {title} {{Disorder-driven itinerant
  quantum criticality of three-dimensional massless Dirac fermions}},}\ }\href
  {\doibase 10.1103/PhysRevB.93.085103} {\bibfield  {journal} {\bibinfo
  {journal} {\prb}\ }\textbf {\bibinfo {volume} {93}},\ \bibinfo {eid} {085103}
  (\bibinfo {year} {2016}{\natexlab{a}})},\ \Eprint
  {http://arxiv.org/abs/1505.07938} {arXiv:1505.07938 [cond-mat.dis-nn]}
  \BibitemShut {NoStop}%
\bibitem [{\citenamefont {{Bera}}\ \emph {et~al.}(2016)\citenamefont {{Bera}},
  \citenamefont {{Sau}},\ and\ \citenamefont {{Roy}}}]{Bera2016}%
  \BibitemOpen
  \bibfield  {author} {\bibinfo {author} {\bibfnamefont {S.}~\bibnamefont
  {{Bera}}}, \bibinfo {author} {\bibfnamefont {J.~D.}\ \bibnamefont {{Sau}}}, \
  and\ \bibinfo {author} {\bibfnamefont {B.}~\bibnamefont {{Roy}}},\ }\bibfield
   {title} {\enquote {\bibinfo {title} {{Dirty Weyl semimetals: Stability,
  phase transition, and quantum criticality}},}\ }\href {\doibase
  10.1103/PhysRevB.93.201302} {\bibfield  {journal} {\bibinfo  {journal}
  {\prb}\ }\textbf {\bibinfo {volume} {93}},\ \bibinfo {eid} {201302} (\bibinfo
  {year} {2016})},\ \Eprint {http://arxiv.org/abs/1507.07551} {arXiv:1507.07551
  [cond-mat.dis-nn]} \BibitemShut {NoStop}%
\bibitem [{\citenamefont {{Roy}}\ \emph {et~al.}(2016)\citenamefont {{Roy}},
  \citenamefont {{Slager}},\ and\ \citenamefont {{Juricic}}}]{Roy2016}%
  \BibitemOpen
  \bibfield  {author} {\bibinfo {author} {\bibfnamefont {B.}~\bibnamefont
  {{Roy}}}, \bibinfo {author} {\bibfnamefont {R.-J.}\ \bibnamefont {{Slager}}},
  \ and\ \bibinfo {author} {\bibfnamefont {V.}~\bibnamefont {{Juricic}}},\
  }\bibfield  {title} {\enquote {\bibinfo {title} {{Global phase diagram of a
  dirty Weyl semimetal}},}\ }\href@noop {} {\bibfield  {journal} {\bibinfo
  {journal} {ArXiv e-prints}\ } (\bibinfo {year} {2016})},\ \Eprint
  {http://arxiv.org/abs/1610.08973} {arXiv:1610.08973 [cond-mat.mes-hall]}
  \BibitemShut {NoStop}%
\bibitem [{\citenamefont {{Burkov}}\ \emph {et~al.}(2011)\citenamefont
  {{Burkov}}, \citenamefont {{Hook}},\ and\ \citenamefont
  {{Balents}}}]{Burkov2011a}%
  \BibitemOpen
  \bibfield  {author} {\bibinfo {author} {\bibfnamefont {A.~A.}\ \bibnamefont
  {{Burkov}}}, \bibinfo {author} {\bibfnamefont {M.~D.}\ \bibnamefont
  {{Hook}}}, \ and\ \bibinfo {author} {\bibfnamefont {L.}~\bibnamefont
  {{Balents}}},\ }\bibfield  {title} {\enquote {\bibinfo {title} {{Topological
  nodal semimetals}},}\ }\href {\doibase 10.1103/PhysRevB.84.235126} {\bibfield
   {journal} {\bibinfo  {journal} {\prb}\ }\textbf {\bibinfo {volume} {84}},\
  \bibinfo {eid} {235126} (\bibinfo {year} {2011})},\ \Eprint
  {http://arxiv.org/abs/1110.1089} {arXiv:1110.1089 [cond-mat.mes-hall]}
  \BibitemShut {NoStop}%
\bibitem [{\citenamefont {{Hosur}}\ \emph {et~al.}(2012)\citenamefont
  {{Hosur}}, \citenamefont {{Parameswaran}},\ and\ \citenamefont
  {{Vishwanath}}}]{Hosur2012}%
  \BibitemOpen
  \bibfield  {author} {\bibinfo {author} {\bibfnamefont {P.}~\bibnamefont
  {{Hosur}}}, \bibinfo {author} {\bibfnamefont {S.~A.}\ \bibnamefont
  {{Parameswaran}}}, \ and\ \bibinfo {author} {\bibfnamefont {A.}~\bibnamefont
  {{Vishwanath}}},\ }\bibfield  {title} {\enquote {\bibinfo {title} {{Charge
  Transport in Weyl Semimetals}},}\ }\href {\doibase
  10.1103/PhysRevLett.108.046602} {\bibfield  {journal} {\bibinfo  {journal}
  {\prl}\ }\textbf {\bibinfo {volume} {108}},\ \bibinfo {eid} {046602}
  (\bibinfo {year} {2012})},\ \Eprint {http://arxiv.org/abs/1109.6330}
  {arXiv:1109.6330 [cond-mat.str-el]} \BibitemShut {NoStop}%
\bibitem [{\citenamefont {{Burkov}}(2015)}]{Burkov2015}%
  \BibitemOpen
  \bibfield  {author} {\bibinfo {author} {\bibfnamefont {A.~A.}\ \bibnamefont
  {{Burkov}}},\ }\bibfield  {title} {\enquote {\bibinfo {title} {{Chiral
  anomaly and transport in Weyl metals}},}\ }\href {\doibase
  10.1088/0953-8984/27/11/113201} {\bibfield  {journal} {\bibinfo  {journal}
  {Journal of Physics Condensed Matter}\ }\textbf {\bibinfo {volume} {27}},\
  \bibinfo {eid} {113201} (\bibinfo {year} {2015})},\ \Eprint
  {http://arxiv.org/abs/1502.07609} {arXiv:1502.07609 [cond-mat.mes-hall]}
  \BibitemShut {NoStop}%
\bibitem [{\citenamefont {{Klier}}\ \emph {et~al.}(2015)\citenamefont
  {{Klier}}, \citenamefont {{Gornyi}},\ and\ \citenamefont
  {{Mirlin}}}]{Klier2015}%
  \BibitemOpen
  \bibfield  {author} {\bibinfo {author} {\bibfnamefont {J.}~\bibnamefont
  {{Klier}}}, \bibinfo {author} {\bibfnamefont {I.~V.}\ \bibnamefont
  {{Gornyi}}}, \ and\ \bibinfo {author} {\bibfnamefont {A.~D.}\ \bibnamefont
  {{Mirlin}}},\ }\bibfield  {title} {\enquote {\bibinfo {title} {{Transversal
  magnetoresistance in Weyl semimetals}},}\ }\href {\doibase
  10.1103/PhysRevB.92.205113} {\bibfield  {journal} {\bibinfo  {journal}
  {\prb}\ }\textbf {\bibinfo {volume} {92}},\ \bibinfo {eid} {205113} (\bibinfo
  {year} {2015})},\ \Eprint {http://arxiv.org/abs/1507.03481} {arXiv:1507.03481
  [cond-mat.mes-hall]} \BibitemShut {NoStop}%
\bibitem [{\citenamefont {{Altland}}\ and\ \citenamefont
  {{Bagrets}}(2015)}]{Altland2015}%
  \BibitemOpen
  \bibfield  {author} {\bibinfo {author} {\bibfnamefont {A.}~\bibnamefont
  {{Altland}}}\ and\ \bibinfo {author} {\bibfnamefont {D.}~\bibnamefont
  {{Bagrets}}},\ }\bibfield  {title} {\enquote {\bibinfo {title} {{Effective
  Field Theory of the Disordered Weyl Semimetal}},}\ }\href {\doibase
  10.1103/PhysRevLett.114.257201} {\bibfield  {journal} {\bibinfo  {journal}
  {\prl}\ }\textbf {\bibinfo {volume} {114}},\ \bibinfo {eid} {257201}
  (\bibinfo {year} {2015})},\ \Eprint {http://arxiv.org/abs/1501.06537}
  {arXiv:1501.06537 [cond-mat.mes-hall]} \BibitemShut {NoStop}%
\bibitem [{\citenamefont {{Ramakrishnan}}\ \emph {et~al.}(2015)\citenamefont
  {{Ramakrishnan}}, \citenamefont {{Milletari}},\ and\ \citenamefont
  {{Adam}}}]{Ramakrishnan2015}%
  \BibitemOpen
  \bibfield  {author} {\bibinfo {author} {\bibfnamefont {N.}~\bibnamefont
  {{Ramakrishnan}}}, \bibinfo {author} {\bibfnamefont {M.}~\bibnamefont
  {{Milletari}}}, \ and\ \bibinfo {author} {\bibfnamefont {S.}~\bibnamefont
  {{Adam}}},\ }\bibfield  {title} {\enquote {\bibinfo {title} {{Transport and
  magnetotransport in three-dimensional Weyl semimetals}},}\ }\href {\doibase
  10.1103/PhysRevB.92.245120} {\bibfield  {journal} {\bibinfo  {journal}
  {\prb}\ }\textbf {\bibinfo {volume} {92}},\ \bibinfo {eid} {245120} (\bibinfo
  {year} {2015})},\ \Eprint {http://arxiv.org/abs/1501.03815} {arXiv:1501.03815
  [cond-mat.mes-hall]} \BibitemShut {NoStop}%
\bibitem [{\citenamefont {{Khalaf}}\ and\ \citenamefont
  {{Ostrovsky}}(2016)}]{Khalaf2016}%
  \BibitemOpen
  \bibfield  {author} {\bibinfo {author} {\bibfnamefont {E.}~\bibnamefont
  {{Khalaf}}}\ and\ \bibinfo {author} {\bibfnamefont {P.~M.}\ \bibnamefont
  {{Ostrovsky}}},\ }\bibfield  {title} {\enquote {\bibinfo {title}
  {{Localization effects on magnetotransport of a disordered Weyl
  semimetal}},}\ }\href@noop {} {\bibfield  {journal} {\bibinfo  {journal}
  {ArXiv e-prints}\ } (\bibinfo {year} {2016})},\ \Eprint
  {http://arxiv.org/abs/1611.09839} {arXiv:1611.09839 [cond-mat.mes-hall]}
  \BibitemShut {NoStop}%
\bibitem [{\citenamefont {{Tabert}}\ \emph {et~al.}(2016)\citenamefont
  {{Tabert}}, \citenamefont {{Carbotte}},\ and\ \citenamefont
  {{Nicol}}}]{Tabert2016}%
  \BibitemOpen
  \bibfield  {author} {\bibinfo {author} {\bibfnamefont {C.~J.}\ \bibnamefont
  {{Tabert}}}, \bibinfo {author} {\bibfnamefont {J.~P.}\ \bibnamefont
  {{Carbotte}}}, \ and\ \bibinfo {author} {\bibfnamefont {E.~J.}\ \bibnamefont
  {{Nicol}}},\ }\bibfield  {title} {\enquote {\bibinfo {title} {{Optical and
  transport properties in three-dimensional Dirac and Weyl semimetals}},}\
  }\href {\doibase 10.1103/PhysRevB.93.085426} {\bibfield  {journal} {\bibinfo
  {journal} {\prb}\ }\textbf {\bibinfo {volume} {93}},\ \bibinfo {eid} {085426}
  (\bibinfo {year} {2016})}\BibitemShut {NoStop}%
\bibitem [{\citenamefont {{Lundgren}}\ \emph {et~al.}(2014)\citenamefont
  {{Lundgren}}, \citenamefont {{Laurell}},\ and\ \citenamefont
  {{Fiete}}}]{Lundgren2014}%
  \BibitemOpen
  \bibfield  {author} {\bibinfo {author} {\bibfnamefont {R.}~\bibnamefont
  {{Lundgren}}}, \bibinfo {author} {\bibfnamefont {P.}~\bibnamefont
  {{Laurell}}}, \ and\ \bibinfo {author} {\bibfnamefont {G.~A.}\ \bibnamefont
  {{Fiete}}},\ }\bibfield  {title} {\enquote {\bibinfo {title} {{Thermoelectric
  properties of Weyl and Dirac semimetals}},}\ }\href {\doibase
  10.1103/PhysRevB.90.165115} {\bibfield  {journal} {\bibinfo  {journal}
  {\prb}\ }\textbf {\bibinfo {volume} {90}},\ \bibinfo {eid} {165115} (\bibinfo
  {year} {2014})},\ \Eprint {http://arxiv.org/abs/1407.1435} {arXiv:1407.1435
  [cond-mat.str-el]} \BibitemShut {NoStop}%
\bibitem [{\citenamefont {{Rosenstein}}\ and\ \citenamefont
  {{Lewkowicz}}(2013)}]{Rosenstein2013}%
  \BibitemOpen
  \bibfield  {author} {\bibinfo {author} {\bibfnamefont {B.}~\bibnamefont
  {{Rosenstein}}}\ and\ \bibinfo {author} {\bibfnamefont {M.}~\bibnamefont
  {{Lewkowicz}}},\ }\bibfield  {title} {\enquote {\bibinfo {title} {{Dynamics
  of electric transport in interacting Weyl semimetals}},}\ }\href {\doibase
  10.1103/PhysRevB.88.045108} {\bibfield  {journal} {\bibinfo  {journal}
  {\prb}\ }\textbf {\bibinfo {volume} {88}},\ \bibinfo {eid} {045108} (\bibinfo
  {year} {2013})},\ \Eprint {http://arxiv.org/abs/1304.7506} {arXiv:1304.7506
  [cond-mat.str-el]} \BibitemShut {NoStop}%
\bibitem [{\citenamefont {{Ashby}}\ and\ \citenamefont
  {{Carbotte}}(2014)}]{Ashby2014}%
  \BibitemOpen
  \bibfield  {author} {\bibinfo {author} {\bibfnamefont {P.~E.~C.}\
  \bibnamefont {{Ashby}}}\ and\ \bibinfo {author} {\bibfnamefont {J.~P.}\
  \bibnamefont {{Carbotte}}},\ }\bibfield  {title} {\enquote {\bibinfo {title}
  {{Chiral anomaly and optical absorption in Weyl semimetals}},}\ }\href
  {\doibase 10.1103/PhysRevB.89.245121} {\bibfield  {journal} {\bibinfo
  {journal} {\prb}\ }\textbf {\bibinfo {volume} {89}},\ \bibinfo {eid} {245121}
  (\bibinfo {year} {2014})},\ \Eprint {http://arxiv.org/abs/1405.7034}
  {arXiv:1405.7034 [cond-mat.mes-hall]} \BibitemShut {NoStop}%
\bibitem [{\citenamefont {{Nandkishore}}\ \emph {et~al.}(2014)\citenamefont
  {{Nandkishore}}, \citenamefont {{Huse}},\ and\ \citenamefont
  {{Sondhi}}}]{Nandkishore2014}%
  \BibitemOpen
  \bibfield  {author} {\bibinfo {author} {\bibfnamefont {R.}~\bibnamefont
  {{Nandkishore}}}, \bibinfo {author} {\bibfnamefont {D.~A.}\ \bibnamefont
  {{Huse}}}, \ and\ \bibinfo {author} {\bibfnamefont {S.~L.}\ \bibnamefont
  {{Sondhi}}},\ }\bibfield  {title} {\enquote {\bibinfo {title} {{Rare region
  effects dominate weakly disordered three-dimensional Dirac points}},}\ }\href
  {\doibase 10.1103/PhysRevB.89.245110} {\bibfield  {journal} {\bibinfo
  {journal} {\prb}\ }\textbf {\bibinfo {volume} {89}},\ \bibinfo {eid} {245110}
  (\bibinfo {year} {2014})},\ \Eprint {http://arxiv.org/abs/1405.2336}
  {arXiv:1405.2336 [cond-mat.stat-mech]} \BibitemShut {NoStop}%
\bibitem [{\citenamefont {{Biswas}}\ and\ \citenamefont
  {{Ryu}}(2014)}]{Biswas2014}%
  \BibitemOpen
  \bibfield  {author} {\bibinfo {author} {\bibfnamefont {R.~R.}\ \bibnamefont
  {{Biswas}}}\ and\ \bibinfo {author} {\bibfnamefont {S.}~\bibnamefont
  {{Ryu}}},\ }\bibfield  {title} {\enquote {\bibinfo {title} {{Diffusive
  transport in Weyl semimetals}},}\ }\href {\doibase
  10.1103/PhysRevB.89.014205} {\bibfield  {journal} {\bibinfo  {journal}
  {\prb}\ }\textbf {\bibinfo {volume} {89}},\ \bibinfo {eid} {014205} (\bibinfo
  {year} {2014})},\ \Eprint {http://arxiv.org/abs/1309.3278} {arXiv:1309.3278
  [cond-mat.mes-hall]} \BibitemShut {NoStop}%
\bibitem [{\citenamefont {{Skinner}}(2014)}]{Skinner2014}%
  \BibitemOpen
  \bibfield  {author} {\bibinfo {author} {\bibfnamefont {B.}~\bibnamefont
  {{Skinner}}},\ }\bibfield  {title} {\enquote {\bibinfo {title} {{Coulomb
  disorder in three-dimensional Dirac systems}},}\ }\href {\doibase
  10.1103/PhysRevB.90.060202} {\bibfield  {journal} {\bibinfo  {journal}
  {\prb}\ }\textbf {\bibinfo {volume} {90}},\ \bibinfo {eid} {060202} (\bibinfo
  {year} {2014})},\ \Eprint {http://arxiv.org/abs/1406.2318} {arXiv:1406.2318
  [cond-mat.dis-nn]} \BibitemShut {NoStop}%
\bibitem [{\citenamefont {{Ominato}}\ and\ \citenamefont
  {{Koshino}}(2015)}]{Ominato2015}%
  \BibitemOpen
  \bibfield  {author} {\bibinfo {author} {\bibfnamefont {Y.}~\bibnamefont
  {{Ominato}}}\ and\ \bibinfo {author} {\bibfnamefont {M.}~\bibnamefont
  {{Koshino}}},\ }\bibfield  {title} {\enquote {\bibinfo {title} {{Quantum
  transport in three-dimensional Weyl electron system in the presence of
  charged impurity scattering}},}\ }\href {\doibase 10.1103/PhysRevB.91.035202}
  {\bibfield  {journal} {\bibinfo  {journal} {\prb}\ }\textbf {\bibinfo
  {volume} {91}},\ \bibinfo {eid} {035202} (\bibinfo {year} {2015})},\ \Eprint
  {http://arxiv.org/abs/1410.0155} {arXiv:1410.0155 [cond-mat.dis-nn]}
  \BibitemShut {NoStop}%
\bibitem [{\citenamefont {{Rodionov}}\ and\ \citenamefont
  {{Syzranov}}(2015)}]{Rodionov2015}%
  \BibitemOpen
  \bibfield  {author} {\bibinfo {author} {\bibfnamefont {Y.~I.}\ \bibnamefont
  {{Rodionov}}}\ and\ \bibinfo {author} {\bibfnamefont {S.~V.}\ \bibnamefont
  {{Syzranov}}},\ }\bibfield  {title} {\enquote {\bibinfo {title}
  {{Conductivity of a Weyl semimetal with donor and acceptor impurities}},}\
  }\href {\doibase 10.1103/PhysRevB.91.195107} {\bibfield  {journal} {\bibinfo
  {journal} {\prb}\ }\textbf {\bibinfo {volume} {91}},\ \bibinfo {eid} {195107}
  (\bibinfo {year} {2015})},\ \Eprint {http://arxiv.org/abs/1503.02078}
  {arXiv:1503.02078 [cond-mat.mes-hall]} \BibitemShut {NoStop}%
\bibitem [{\citenamefont {{Pixley}}\ \emph
  {et~al.}(2016{\natexlab{b}})\citenamefont {{Pixley}}, \citenamefont
  {{Huse}},\ and\ \citenamefont {{Das Sarma}}}]{Pixley2016a}%
  \BibitemOpen
  \bibfield  {author} {\bibinfo {author} {\bibfnamefont {J.~H.}\ \bibnamefont
  {{Pixley}}}, \bibinfo {author} {\bibfnamefont {D.~A.}\ \bibnamefont
  {{Huse}}}, \ and\ \bibinfo {author} {\bibfnamefont {S.}~\bibnamefont {{Das
  Sarma}}},\ }\bibfield  {title} {\enquote {\bibinfo {title}
  {{Rare-Region-Induced Avoided Quantum Criticality in Disordered
  Three-Dimensional Dirac and Weyl Semimetals}},}\ }\href {\doibase
  10.1103/PhysRevX.6.021042} {\bibfield  {journal} {\bibinfo  {journal} {\prx}\
  }\textbf {\bibinfo {volume} {6}},\ \bibinfo {eid} {021042} (\bibinfo {year}
  {2016}{\natexlab{b}})},\ \Eprint {http://arxiv.org/abs/1602.02742}
  {arXiv:1602.02742 [cond-mat.dis-nn]} \BibitemShut {NoStop}%
\bibitem [{\citenamefont {{Ostrovsky}}\ \emph {et~al.}(2006)\citenamefont
  {{Ostrovsky}}, \citenamefont {{Gornyi}},\ and\ \citenamefont
  {{Mirlin}}}]{Ostrovsky2006}%
  \BibitemOpen
  \bibfield  {author} {\bibinfo {author} {\bibfnamefont {P.~M.}\ \bibnamefont
  {{Ostrovsky}}}, \bibinfo {author} {\bibfnamefont {I.~V.}\ \bibnamefont
  {{Gornyi}}}, \ and\ \bibinfo {author} {\bibfnamefont {A.~D.}\ \bibnamefont
  {{Mirlin}}},\ }\bibfield  {title} {\enquote {\bibinfo {title} {{Electron
  transport in disordered graphene}},}\ }\href {\doibase
  10.1103/PhysRevB.74.235443} {\bibfield  {journal} {\bibinfo  {journal}
  {\prb}\ }\textbf {\bibinfo {volume} {74}},\ \bibinfo {eid} {235443} (\bibinfo
  {year} {2006})},\ \Eprint {http://arxiv.org/abs/cond-mat/0609617}
  {cond-mat/0609617} \BibitemShut {NoStop}%
\bibitem [{\citenamefont {Mahan}(1990)}]{Mahan1990}%
  \BibitemOpen
  \bibfield  {author} {\bibinfo {author} {\bibfnamefont {G.D.}\ \bibnamefont
  {Mahan}},\ }\href@noop {} {\emph {\bibinfo {title} {Many-Particle Physics}}}\
  (\bibinfo  {publisher} {Springer},\ \bibinfo {year} {1990})\BibitemShut
  {NoStop}%
\bibitem [{\citenamefont {Sakurai}\ and\ \citenamefont
  {Napolitano}(2011)}]{Sakurai2011}%
  \BibitemOpen
  \bibfield  {author} {\bibinfo {author} {\bibfnamefont {J.J.}\ \bibnamefont
  {Sakurai}}\ and\ \bibinfo {author} {\bibfnamefont {J.}~\bibnamefont
  {Napolitano}},\ }\href@noop {} {\emph {\bibinfo {title} {Modern Quantum
  Mechanics}}}\ (\bibinfo  {publisher} {Addison-Wesley},\ \bibinfo {year}
  {2011})\BibitemShut {NoStop}%
\end{thebibliography}
\end{document}